\documentclass[twocolumn,floatfix,showpacs,superscriptaddress]{revtex4-2}

\usepackage{amssymb,amsmath,amsfonts}
\usepackage{graphicx,graphics}
\usepackage{epsfig}
\usepackage{epstopdf}
\usepackage{amsfonts}
\usepackage{bm,bbm}
\usepackage{amssymb,amsmath,amsfonts}
\usepackage{mathrsfs}
\usepackage{calc}
\usepackage{epsfig}
\usepackage{float}
\usepackage{color}
\usepackage{epstopdf}
\usepackage{bm,amssymb,amsmath}
\usepackage{graphicx,color}

\usepackage{ulem}

\begin{document}

\title{The fluctuation-dissipation relation holds for a macroscopic tracer in an active bath}
\author{Dima Boriskovsky}
\affiliation{Raymond \& Beverly Sackler School of Physics and Astronomy, Tel Aviv University, Tel Aviv 6997801, Israel}
\author{Benjamin Lindner}
\affiliation{Bernstein Center for Computational Neuroscience Berlin, Philippstr.~13, Haus 2, 10115 Berlin, Germany}
\affiliation{Physics Department of Humboldt University Berlin, Newtonstr.~15, 12489 Berlin, Germany}
\author{Yael Roichman}
\email{roichman@tauex.tau.ac.il}
\affiliation{Raymond \& Beverly Sackler School of Physics and Astronomy, Tel Aviv University, Tel Aviv 6997801, Israel}
\affiliation{Raymond \& Beverly Sackler School of Chemistry, Tel Aviv University, Tel Aviv 6997801, Israel}

\date{\today}
\begin{abstract}
The fluctuation-dissipation relation (FDR) links thermal fluctuations and dissipation at thermal equilibrium through temperature. Extending it beyond equilibrium conditions in pursuit of broadening thermodynamics is often feasible, albeit with system-dependent specific conditions. We demonstrate experimentally that a generalized FDR holds for a harmonically trapped tracer colliding with self-propelled walkers. The generalized FDR remains valid across a large spectrum of active fluctuation frequencies, extending from underdamped to critically damped dynamics, which we attribute to a single primary channel for energy input and dissipation in our system.
\end{abstract}

\maketitle



The fluctuation-dissipation relation (FDR) is a fundamental result in nonequilibrium statistical mechanics. Its significance lies in providing a means to compute how a system statistically responds to small external perturbations, in terms of the correlations in the unperturbed dynamics. A classic example of the FDR is the Einstein relation \cite{einstein1905}, $D=\mu k_BT$, linking the diffusivity (fluctuations) $D$ with the mobility (response) $\mu$ for a Brownian particle in equilibrium with a bath at temperature $T$ ($k_B$ denotes the Boltzmann constant). This relation, known as a \textit{static} FDR, is established under constant force conditions. On the other hand, the \textit{dynamic} FDR broadens this concept by encompassing time or frequency-dependent forces via linear response theory \cite{kubo1966,kubo1986}.

Specifically, for a system slightly perturbed from its stable equilibrium condition, the mean response of an observable $x$, denoted as $R_x(t>0)$, to an abrupt arrest of a force $F_0$ at time $t=0$, is connected to the autocorrelation function $C_x(t)$ in the unperturbed condition. The dynamic FDR then states \cite{marconi2008,baldovin2022review}, 
\begin{align}
    R_x(t) = \frac{F_0}{k_BT} C_x(t) . \label{eq:dFDR}
\end{align}
A static FDR is recovered at $t=0$. Namely, $C_x(t)\rightarrow\langle \Delta x^2\rangle_0$ and  $R_x(t)\rightarrow \delta \langle x(0)\rangle$ and we obtain a relation in terms of energies, $F_0\langle \Delta x^2\rangle_0/\delta \langle x(0)\rangle=k_BT$. The dynamic FDR (Eq. \ref{eq:dFDR} and its equivalent forms) has been utilized as a model-free method to assess equilibrium in various experimental systems \cite{martin2001,mizuno2007,turlier2016}. If the relationship in Eq. \ref{eq:dFDR} is violated, it indicates that the system is operating outside of thermal equilibrium.

Extensions of the FDR to perturbations of nonequilibrium steady states were derived previously, both for weak perturbations (e.g., \cite{Berthier2002,prost2009,mehl2010,willareth2017,Lindner2022,PerGut23}) and for strong perturbations leading to a nonlinear form of FDR \cite{baldovin2022review,lippiello2008,engbring2023}. However, most of these relations usually require a variable transformation or are derived for special non-equilibrium models (such as spiking neurons as in \cite{Lindner2022}). Alternatively, in nonequilibrium scenarios, a key theoretical use of the FDR is the introduction of an effective temperature $T_{\text{eff}}$ \cite{ojha2004,villamaina2008,abate2008,maggi2014,szamel2014,dieterich2015,ginot2015,steffenoni2016,maggi2017,han2017,petrelli2020,flenner2020,solon2022,boudet2022,cugliandolo2011}, satisfying a \textit{generalized} FDR by substituting $T=T_{\text{eff}}$ in Eq. \ref{eq:dFDR}. When $T_{\text{eff}}$ remains constant, irrespective of time and perturbation magnitude, the analogy to an equilibrium description becomes rather transparent: the response to an external perturbation is directly proportional to the unperturbed correlations, with $F_0/k_BT_{\text{eff}}$ serving as the constant of proportionality.

An archetypal model for studying the validity range of the FDR out of equilibrium, and the emergence of an effective temperature, is a tracer particle trapped in a harmonic potential and subject to both thermal and active collisions \cite{uhlenbeck1930,bechinger2016review,fodor2016,demery2019,ye2020,shea2022,jayaram2023}, e.g., an optically trapped colloidal particle in a suspension of self-propelled organisms \cite{wu2000,angelani2009,palacci2010,park2020,goerlich2022,paul2022}. 
Within this model, which is typically overdamped, there are two timescales of interest: 
the typical relaxation time $\tau_r$ to the steady state, and the characteristic time of the active stochastic forces $\tau_c$. 
In previous demonstrations involving such a system, it was established that the generalized FDR holds for large times only under the condition that $\tau_c$ is smaller than the longest relaxation time in the steady state \cite{dieterich2015}. Specifically, the FDR holds when the external driving dominantly determines both fluctuations and dissipation. 

On a different scale, far from thermal equilibrium, numerical investigations of a tracer particle within an athermal uniformly driven granular gas \cite{puglisi1999,poschel2001,van2004} validate a generalized FDR solely for nearly elastic collisions \cite{puglisi2002,shokef2004}. In this scenario, the effective temperature was determined through the tracer's mean kinetic energy, $T_{\text{eff}}\sim \langle v^2 \rangle$ \cite{villamaina2008}. 
We note that the effective temperature defined through the FDR was shown to be equivalent to other independent definitions of effective temperatures in driven granular media \cite{chastaing2017,zeng2022}.

These findings suggest that the generalized FDR remains applicable when the same physical process governs the dissipation of external perturbations while simultaneously driving the tracer's fluctuations around its steady state. 
To test this hypothesis for a completely athermal system,  we study a harmonically trapped macroscopic tracer particle in an active bath of self-propelled walkers. 
In our previous work \cite{engbring2023}, we demonstrated that the tracer particle's dynamics are consistent with Markovian dynamics, as evidenced by the fulfillment of a non-linear FDR. 
Notably, this non-linear FDR applies only to a unique conjugated variable and does not yield a straightforward definition of an effective temperature. 
Here, we find that a generalized FDR holds in a wide range of experimental conditions with a naturally defined effective temperature that coincides with the unperturbed mean potential energy of the tracer, $\frac{1}{2}k_BT_{\text{eff}}= \frac{1}{2} k\langle \Delta x^2\rangle_0$. 
Here, $\langle \Delta x^2\rangle_0=\langle (x-\langle x\rangle)^2\rangle$ is the variance of the tracer's position in the unperturbed steady state.


\begin{figure}[t] 
    \centering
    \includegraphics[width=0.23\textwidth]{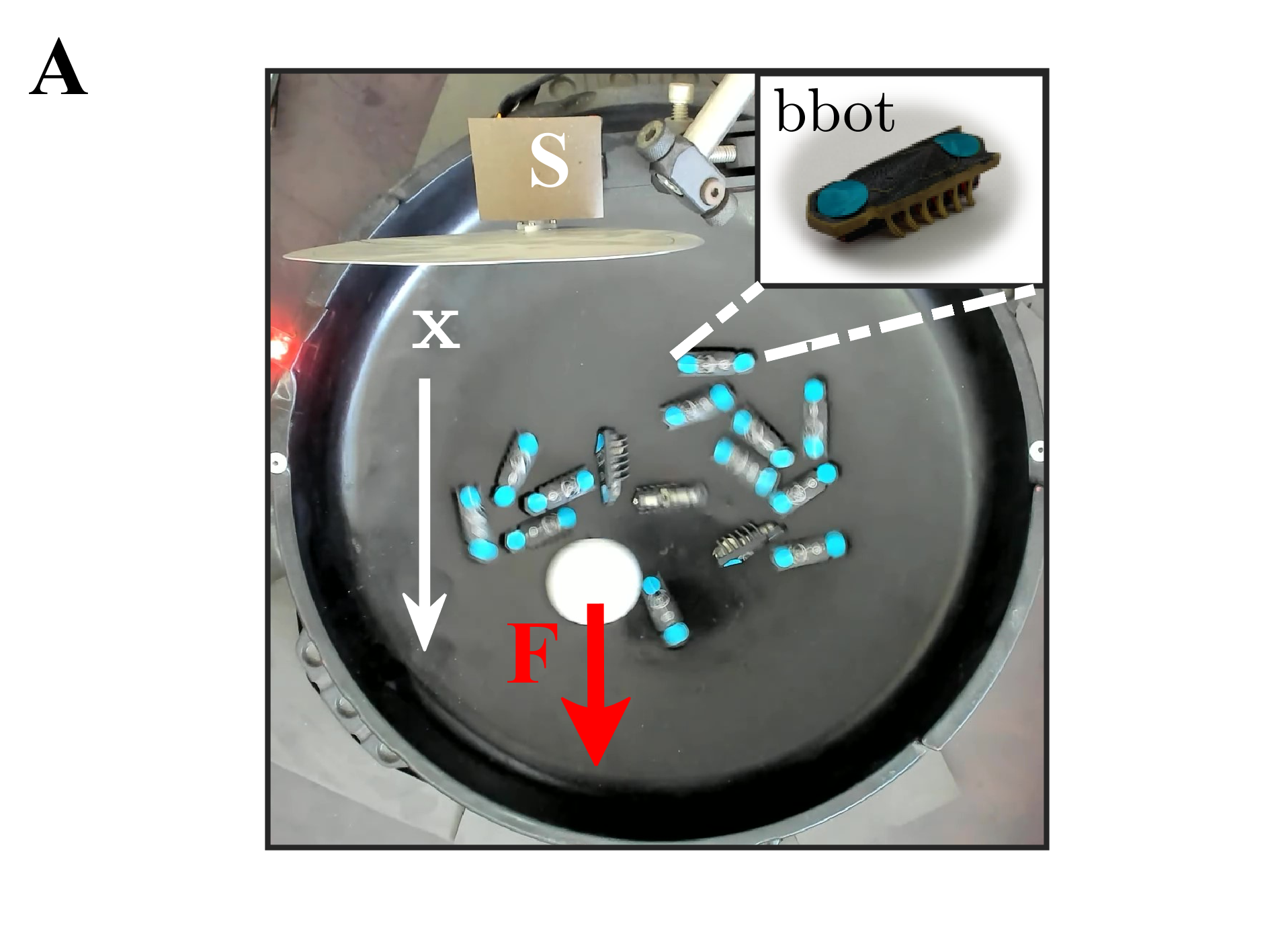}
    \includegraphics[width=0.23\textwidth]{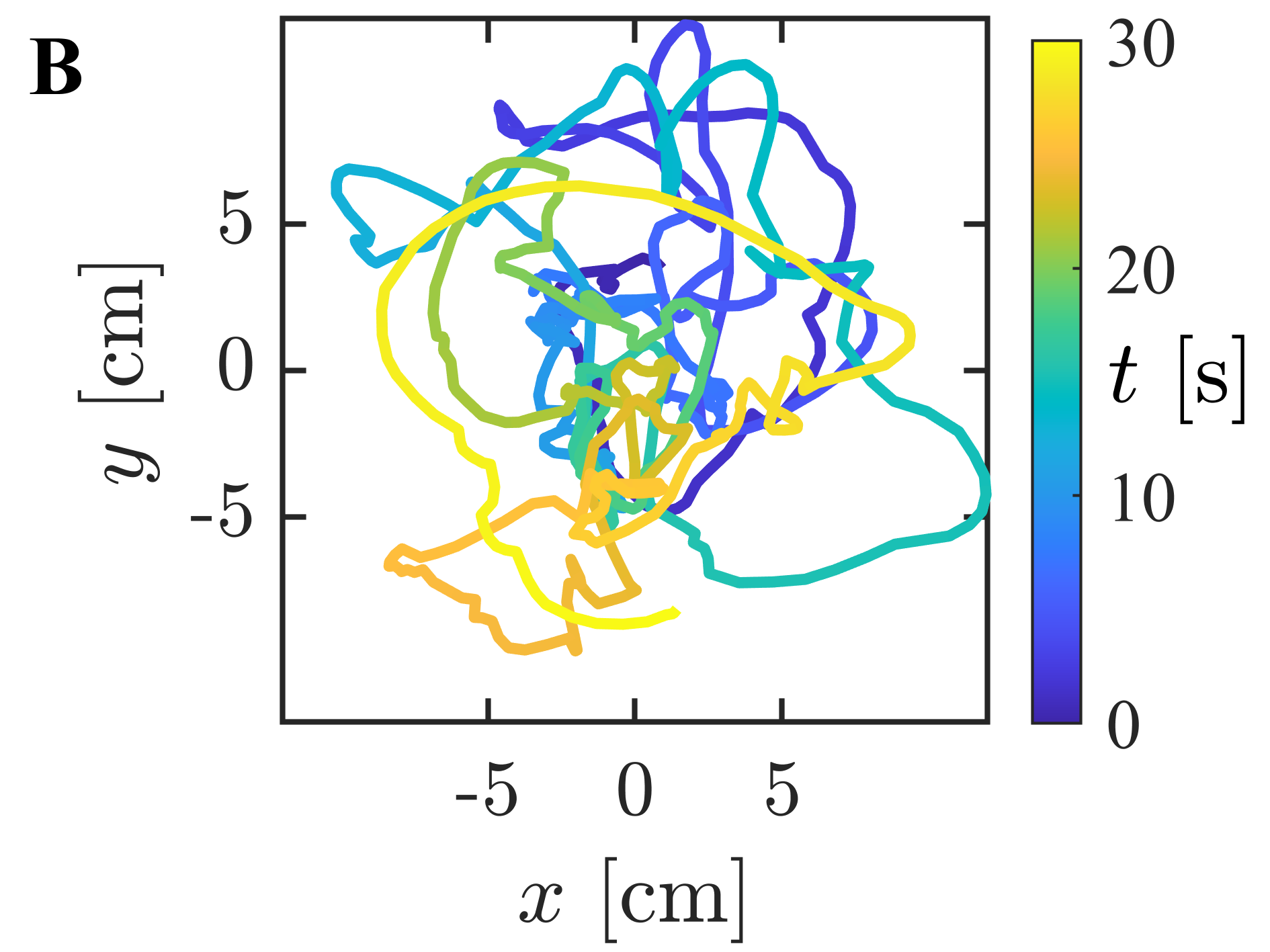}
    \includegraphics[width=0.23\textwidth]{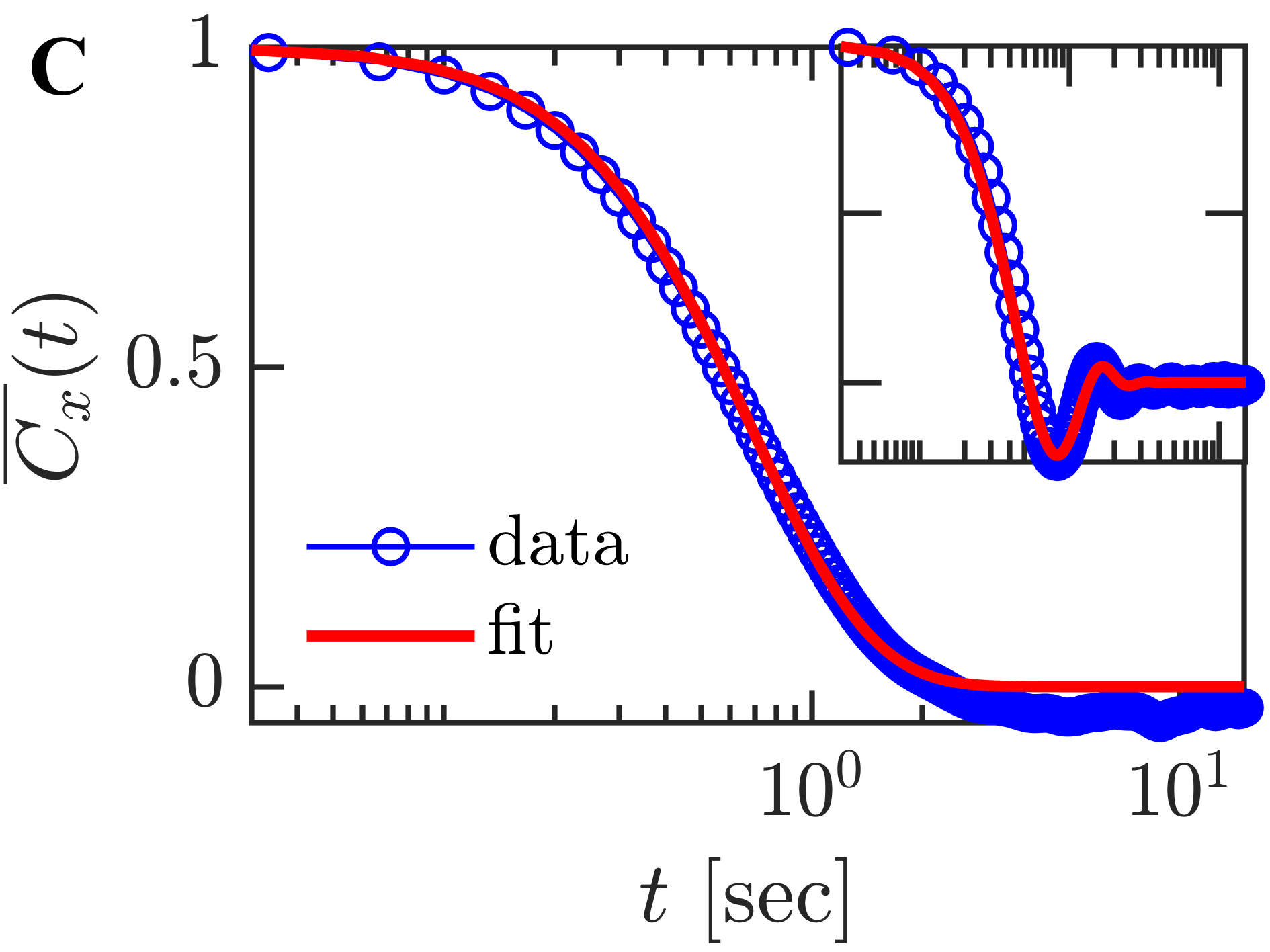}
    \includegraphics[width=0.23\textwidth]{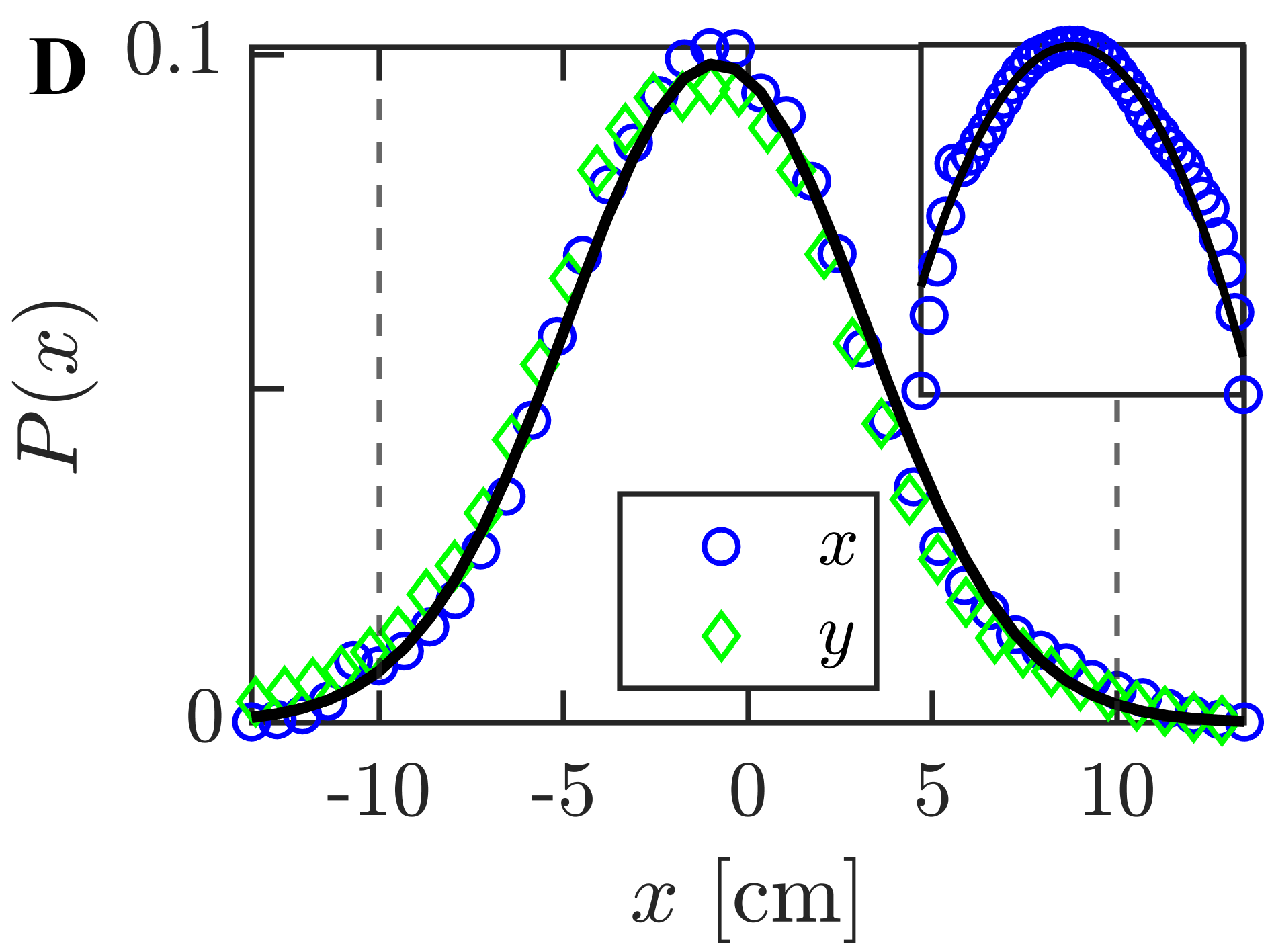}
    \caption{ 
    \textbf{Experimental setup.} \ 
    \textbf{A.} \ A Styrofoam ball (diameter $\sim4$~cm) is trapped in a gravitational harmonic potential, a plastic bowl (diameter 38cm, depth 5cm), and subjected to collisions with $N_b=15$ self-propelled bbots (inset: standard bbot, $4\times1$~cm). The ball is repeatedly perturbed with a uniform air stream created by an external fan along the $x$-axis (white arrow) to test a fluctuation-response relation. To enforce an abrupt onset and release of the perturbation, a mechanical shutter is used (denoted by 'S'). 
    \textbf{B.} \ Part ($30$~s) of a typical trajectory of the Styrofoam ball in the bbot arena.
    \textbf{C.} \ Normalized measured position autocorrelation as a function of time (circles), and fit to Eq. \ref{eq:PCF}, with $\tau_r=0.336\approx\tau_\Omega=0.340$~s (solid line). 
    The inset shows the autocorrelation of an equivalent experiment with $N_b=4$ ($\tau_r=0.54$~s, $\tau_\Omega=0.23$~s).
    The results present an average over $M=375$ sequences.
    \textbf{D.} \ The cross-section probability density distributions across the $x$ axis (blue), and the $y$ axis (green), for $N_b=15$.
    }
    \label{fig:1}
\end{figure}

Our experiments are conducted as depicted in Fig.~\ref{fig:1}\textbf{A}: a Styrofoam ball is placed in a parabolic plastic arena and is driven into a nonequilibrium steady state by random collisions with an assembly of vibration-driven bristle robots (bbots, Hexbugs$^{\text{TM}}$ nano, \cite{giomi2013,dauchot2019,altshuler2023,engbring2023,chor2023}). 
The motion of the ball is recorded by a top-view camera (Brio 4K, Logitech) at $30~\text{fps}$. 
An image analysis algorithm is used to extract the ball's trajectory projection on the XY plane from top-view images. We assume independent motion along all three axes and focus on analyzing the tracer's movement along the X-axis, where the perturbation occurs.
The collisions are non-elastic, i.e., the tracer's restitution coefficient depends on impact velocities ($r\approx 0.8$ for low velocities, see SI$^{\dag}$). 
We note that in contrast to shaken granular matter, the embedding media is active on the single particle level and exhibits emergent collective motion typical for active matter systems \cite{giomi2013,chor2023}. 
The characteristics of the bbot bath are added in the SI$^{\dag}$.
Namely, a single bbot generally performs circular motion in a parabolic arena when it has sufficient inertia \cite{dauchot2019}.
Placing additional bbots and a tracer ball in the harmonic well randomizes the bbot motion due to collisions, resulting in an active gas-like state (see SI movie 1).
Notably, the speed distribution of the bbots remains independent of $N_b$, while their spatial distribution increasingly deviates from the initial circular motion as $N_b$ increases.
As a result, the tracer is randomly kicked by the bbots and exhibits Brownian-like motion (Fig.~\ref{fig:1}\textbf{B}).

To measure the system's mean response, we introduce a mechanical perturbation by activating an external fan (Yate Loon electronics $12$~V $0.30$~A cooling fan) to create an airflow. After the tracer settles into a perturbed steady state, we abruptly deactivate the disturbance, with a physical shutter, allowing the ball to return to its original steady state.
We use the number of bbots $N_b$ as a control parameter, keeping the trapping stiffness constant, i.e., the arena shape and tracer size are held constant. 
Naturally, the mean free time between collisions $\tau_{c}$ decreases with $N_b$, and the collision frequency increases \footnote[3]{The mean free time between collisions $\tau_c$ was estimated by tracking both the bbots and the tracer, averaging over the times in which there's no physical contact between the bbots and the tracer.} (see below). 

Since thermal forces are negligible, the ensemble of bbots $N_b$ acts as a \textit{dry} active bath, meaning that the tracer is subjected to a single active noise source (bbot impacts).
Therefore, the noise characteristic time $\tau_c$ and the stationary relaxation time $\tau_r$ are intrinsic system timescales. 
Particularly, we expect that sufficiently frequent collisions will dominate the tracer's dissipation rate of small external perturbations. 
In this scenario, the main channel for dissipation is the same process that induces active fluctuations, and we expect a generalized FDR to hold as was suggested by Kubo for equilibrium conditions \cite{kubo1966}. 


\textit{Stationary dynamics and statistics.--}
We start our experiments by characterizing the nonequilibrium steady state, which we will later perturb. To this end, we focus on a system with a relatively large number of bbots,  $N_b=15$, in which the tracer particle is subjected to frequent collisions with $\tau_c=0.28$~s, extracted by image analysis identification of collisions. 
We record 375 experiments of duration $30$~s from which the tracer particle trajectory is extracted. From the extracted trajectories, we calculate the tracer's (ensemble) average normalized position autocorrelation, $\overline{C_x}(t)$ (Fig.~\ref{fig:1}\textbf{C}).
It turns out that this correlation is well fit by the generic  autocorrelation function of a damped noisy harmonic oscillator \cite{uhlenbeck1930,shokef2004},
\begin{align} \label{eq:PCF}
    \overline{C_x}(t)\equiv \frac{C_x(t)}{\langle \Delta x^2\rangle_0} = e^{-t/\tau_r}\left( \cos\omega t+\frac{\sin\omega t}{\tau_r \omega} \right),
\end{align}  
where $\langle \Delta x^2\rangle_0=C_x(0)$ is the variance and $\omega^2=\tau_\Omega^{-2}-\tau_r^{-2}$. This fit provides effective values of the system's time scale, the damping rate, $2\tau_r^{-1}$,  and the (effective) natural frequency of the harmonic trap \footnote[4]{Note that since the bbots are confined within the same gravitational well as the tracer, the harmonic trap's natural frequency $\tau_\Omega^{-1}$ depends on $N_b$ and does not coincide with the measured stiffness, i.e., $\sqrt{k/m}$.}, $\tau_\Omega^{-1}$, for which we obtain $\tau_r=0.336\approx\tau_\Omega=0.340$~s. 
These values suggest that the tracer dynamics exhibit critically damped behavior.
We note that underdamped relaxation ($\tau_r>\tau_\Omega$) is observed for much lower $N_b=4$ (inset of Fig.~\ref{fig:1}\textbf{C}).

\begin{figure}[t] 
    \centering
    \includegraphics[width=0.23\textwidth]{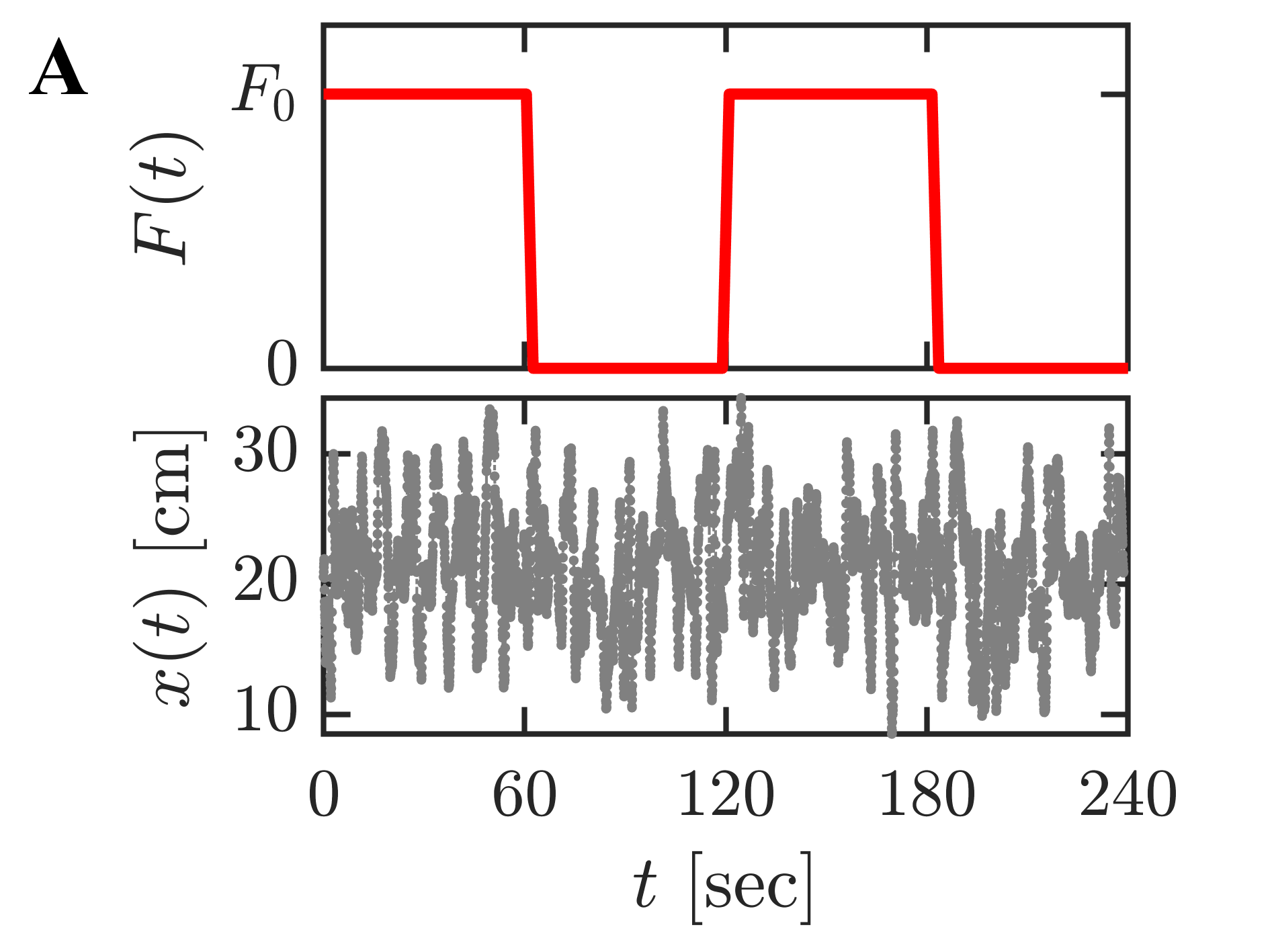}
    \includegraphics[width=0.23\textwidth]{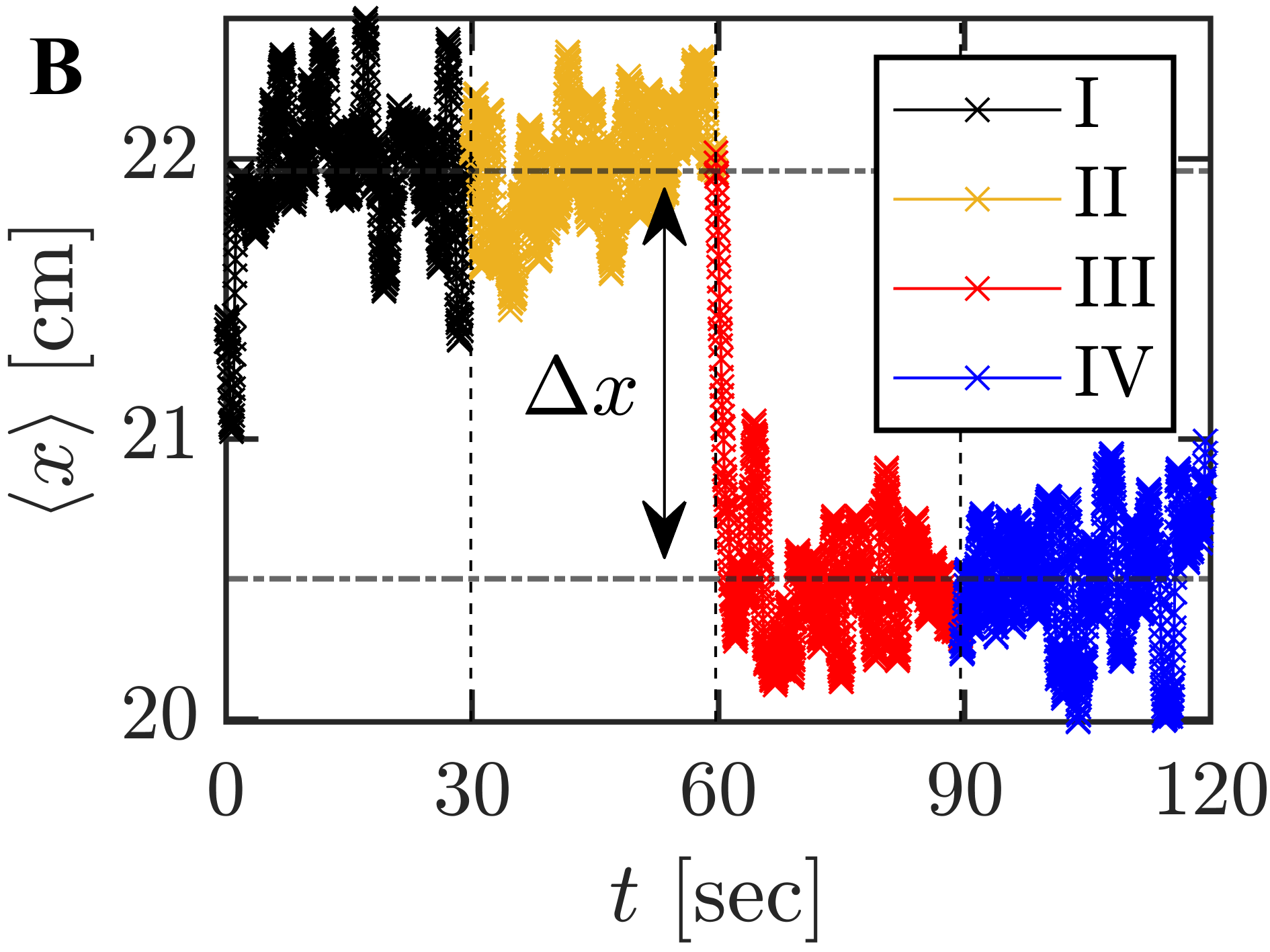}
    \includegraphics[width=0.23\textwidth]{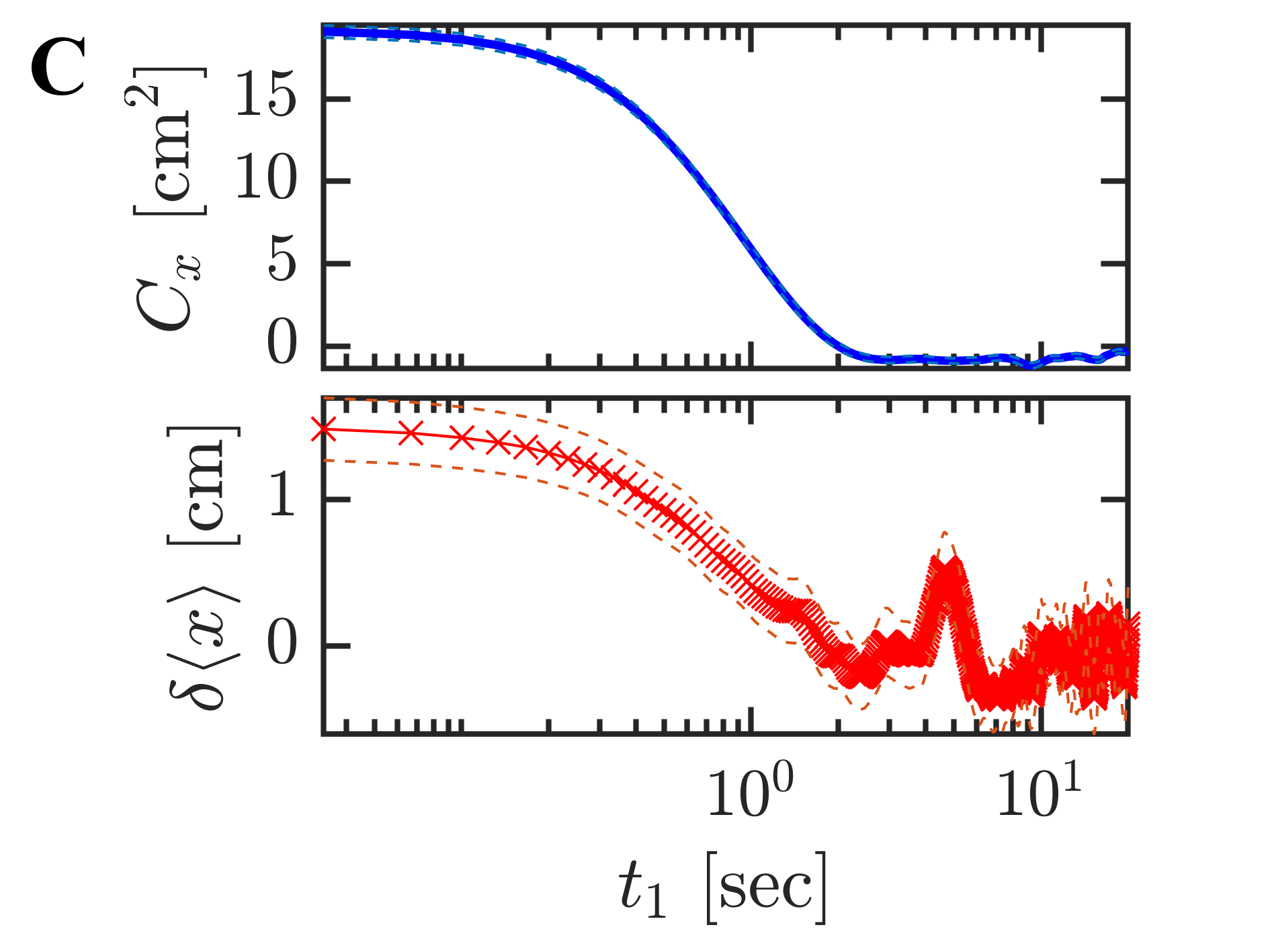}
    \includegraphics[width=0.23\textwidth]{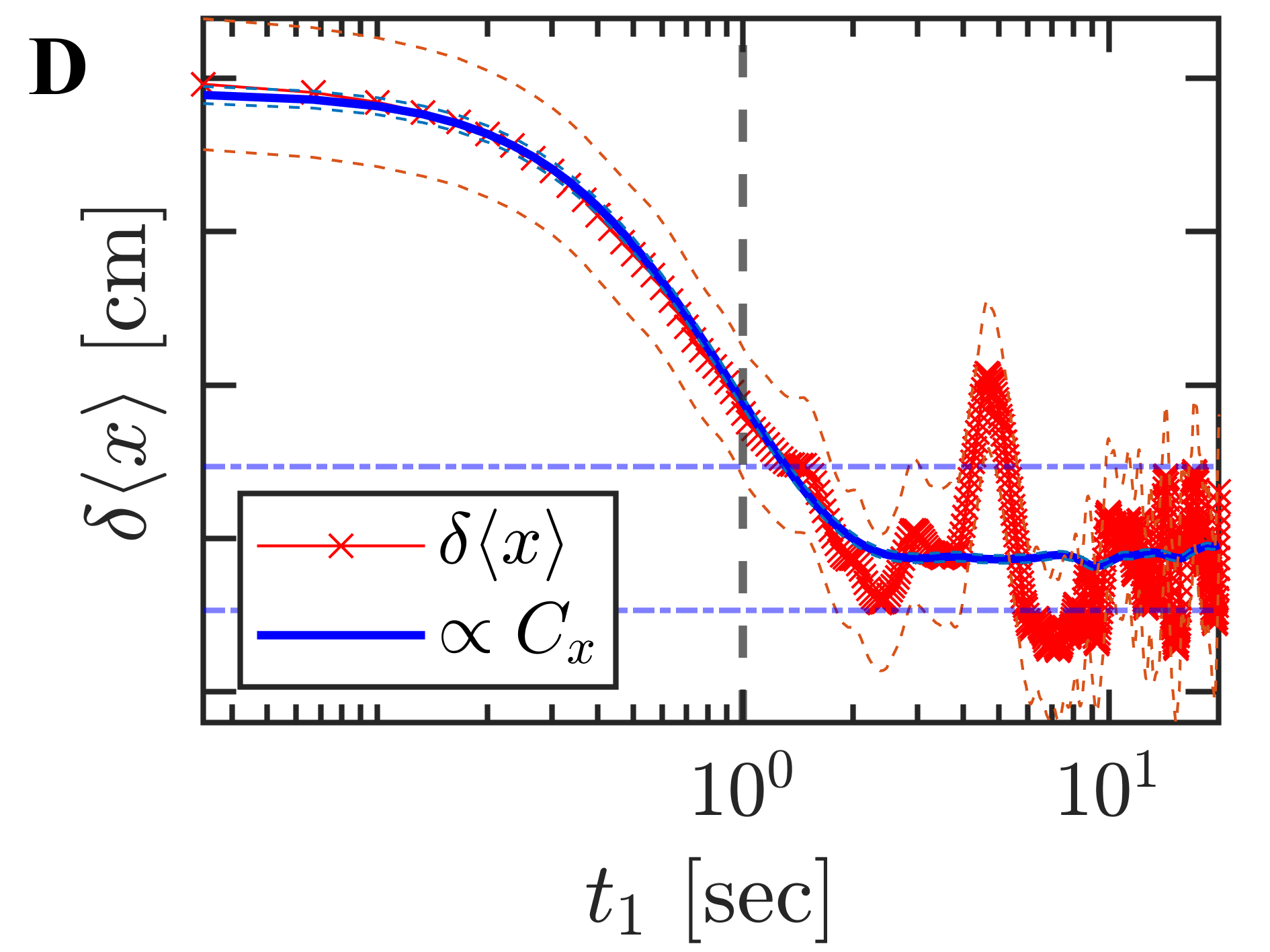}
    \caption{
    \textbf{An experimental FDR test.} \textbf{A.} \ Typical perturbation sequence $x(t)$: the fan is turned on for a minute and then abruptly turned off for the following minute.
    \textbf{B.} \ The time-dependent mean value $\langle x(t)\rangle$ is obtained within a time-window of $2$ minutes. 
    The displacement $\Delta x$ is the mean distance between states II and IV. 
    \textbf{C.} \ The individual time-dependent quantities of transient response $\langle x(t_1)\rangle - \langle x\rangle_0$ (bottom), and unperturbed autocorrelation function $C_x(t_1)$ (top), are displayed. The dashed lines are standard errors.
    \textbf{D.} \ The generalized FDR of Eq. \ref{eq:dFDR} with an effective temperature $T=T_{\text{eff}}\sim \langle \Delta x^2\rangle_0$.
    The results present an average over $M=375$ sequences with $N_b=15$ and fan operating voltage $10$V.
    The vertical dashed line is $T_c=1$~s and the horizontal dot-dashed lines are the standard deviations of part IV.
    }
    \label{fig:2}
\end{figure}

Next, we consider the position probability distribution $P(x,y)$, of the tracer particle (Fig.~\ref{fig:1}\textbf{D}).
We find that $P(x,y)$ fits well to a Boltzmann-like Gaussian distribution in both $x$ and $y$ projections, with a small expected deviation near the arena boundaries ($|x|>10$~cm). We estimate the gravitational potential acting on the tracer particle by $U_x(x)=\frac{1}{2}kx^2$, with $k=mga=28.2 \pm 3~\text{g/s}^{-2}$, $m=1 \pm 0.1$~g - the tracer mass, $g$ - the gravitational acceleration (not to be confused with the unit of mass, g, used before), and $a$ - the curvature of the arena. Plotting the estimated $U_x(x)$ and fitting it to our Boltzmann-like position distribution (see SI$^{\dag}$), we obtain an effective temperature of $T_\text{eff}\approx 3.1\cdot 10^{18}$~K, 16 orders of magnitude higher than room temperature. 
Interestingly, the tracer’s potential energy set by the effective temperature is comparable to the mean kinetic energy of the bristle bots, $E_b \approx 7\mu J$: we observe that $k_B T_\text{eff} \approx 7· E_b$.


\textit{Fluctuation-dissipation relation.--}
With a full characterization of the nonequilibrium steady state dynamics and statistics of the system, we continue by measuring the system's response to a small mechanical perturbation ($10$V fan operating voltage, force $F_0\approx 62~\mu\text{N}$). We use the following protocol to measure the ensemble average response of the system (see also Ref.~\cite{engbring2023} for a similar procedure): during an experiment, every two minutes,  the fan is turned on for a minute and abruptly turned off at $t_0=60$~s for the following minute (see Fig. \ref{fig:2}\textbf{A}). 
Fig. \ref{fig:2}\textbf{B} shows the time-dependent mean value $\langle x(t)\rangle$ for $M=375$ perturbation sequences, constituting both the mean motion of the tracer at the perturbed and unperturbed steady states (parts II and IV) and the transition between them (parts I and III). 

The response of the system to the perturbation arrest, at ${t_1}=t-t_0$, given by $R_x(t_1)=\delta\langle x(t_1)\rangle=\langle x(t_1)\rangle - \langle x\rangle_0$, and the unperturbed autocorrelation function, $C_{x}(t_1)=\langle x(t_1)x(0)\rangle_0$, are therefore measured from parts III, and IV respectively (Fig.~\ref{fig:2}\textbf{C}). According to the generalized FDR, Eq.~\ref{eq:dFDR} should hold exactly for our measured response and autocorrelation if we define $T=T_\text{eff}$ and  $F_0=k\Delta x$, where $\Delta x\equiv\langle x\rangle_{F_0}-\langle x \rangle_0$. This is verified in Fig.~\ref{fig:2}\textbf{D} with good accuracy, where we have used the value of $T_\text{eff}$ obtained from the stationary position distribution. 
We note that at $t_1>T_c= 1$~s the stationary noise becomes significant and the agreement or disagreement of $C_x$ and $R_x$ is hard to assess. Therefore, we cannot rule out an FDR violation beyond $T_c$.
Since $T_\text{eff}$ fulfills the generalized FDR for any $t_1<T_c$, it also fulfills it at $t_1=0$.  Therefore, $k_BT_{\text{eff}}= k\langle \Delta x^2\rangle_0$, which is trivial for a system exhibiting Boltzmann-like statistics. However, this relation should generally hold for any $P(x,y)$ (with different $N_b$).

Our experimental findings establish a consistent effective temperature, $T_\text{eff}$, derived from the generalized FDR. This $T_\text{eff}$ is consistent with the temperature determined by the average potential energy and remains independent of the force (see SI$^{\dag}$). With this defined $T_\text{eff}$, we can further explore the conditions under which the generalized FDR remains applicable.


\begin{figure}[t] 
    \centering
    \includegraphics[width=0.23\textwidth]{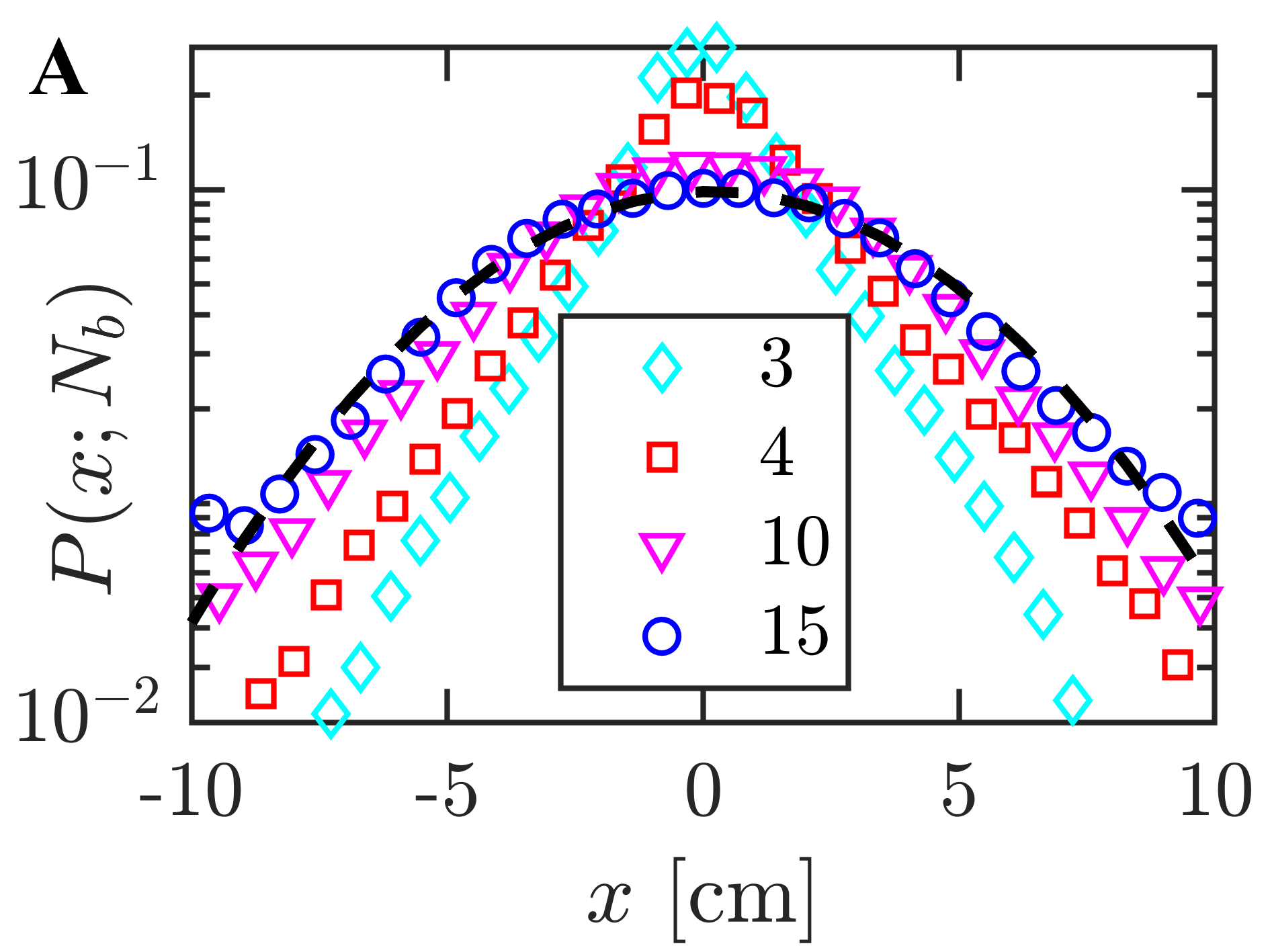}
    \includegraphics[width=0.23\textwidth]{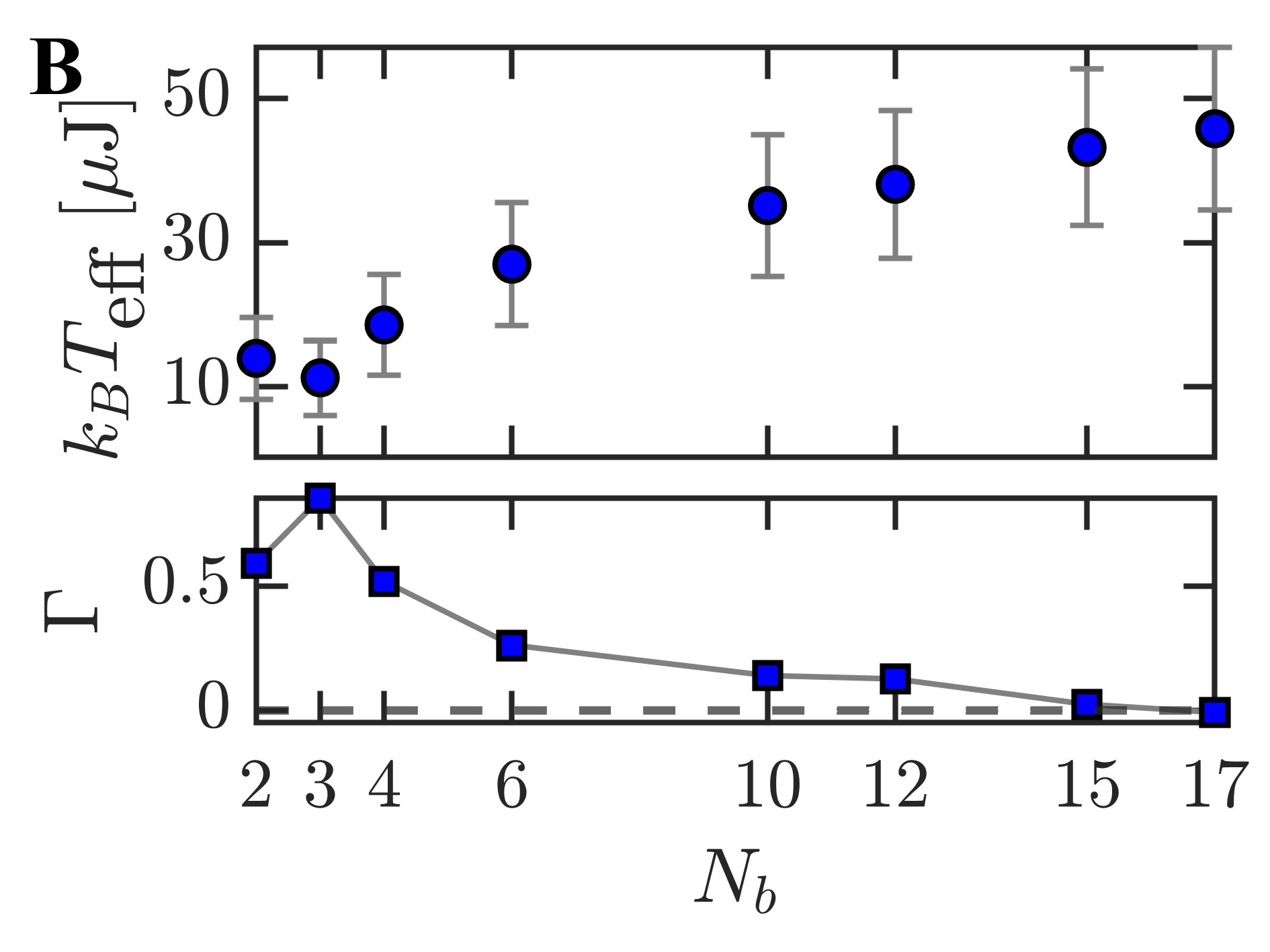}
    \includegraphics[width=0.23\textwidth]{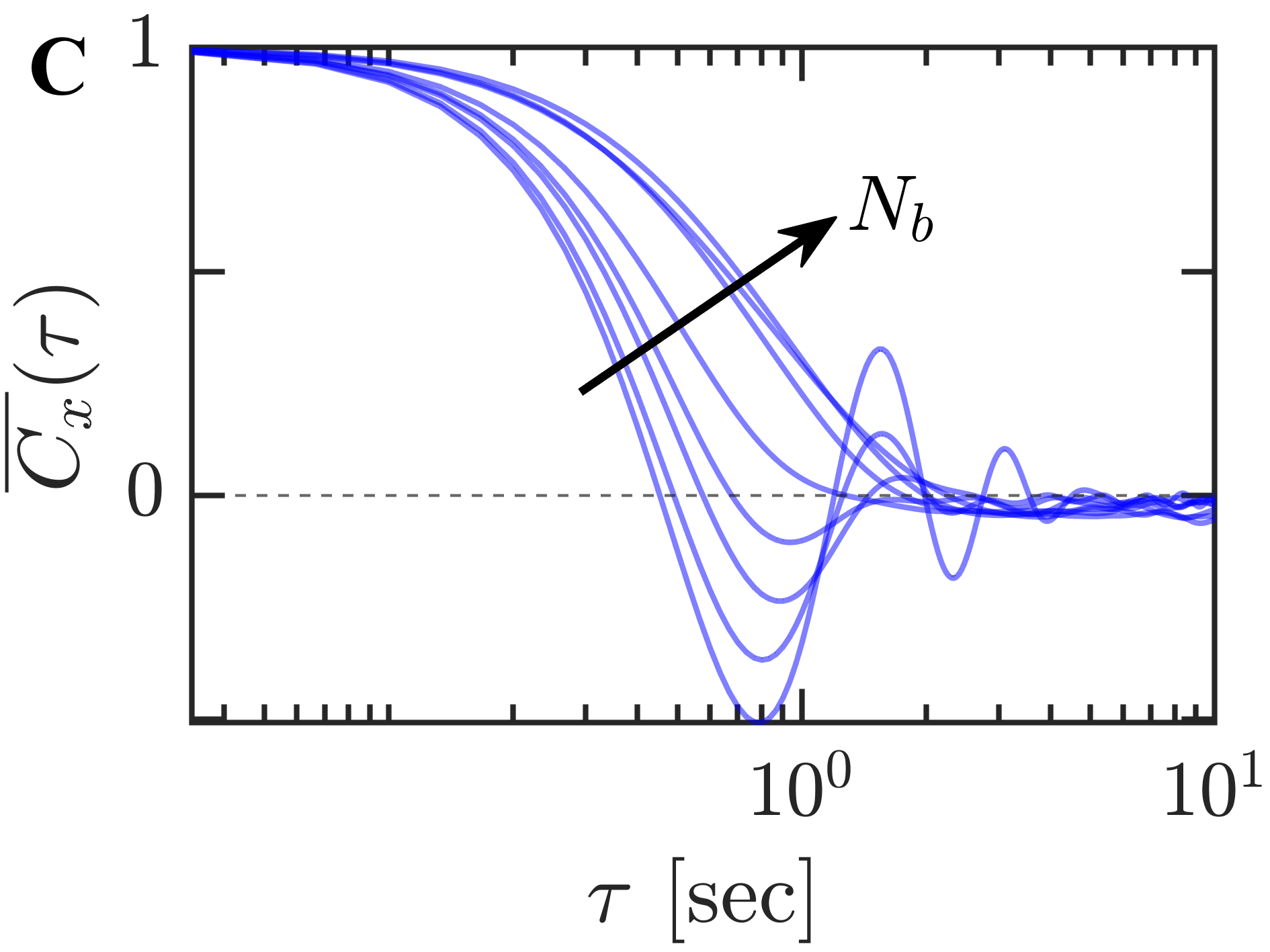}
    \includegraphics[width=0.23\textwidth]{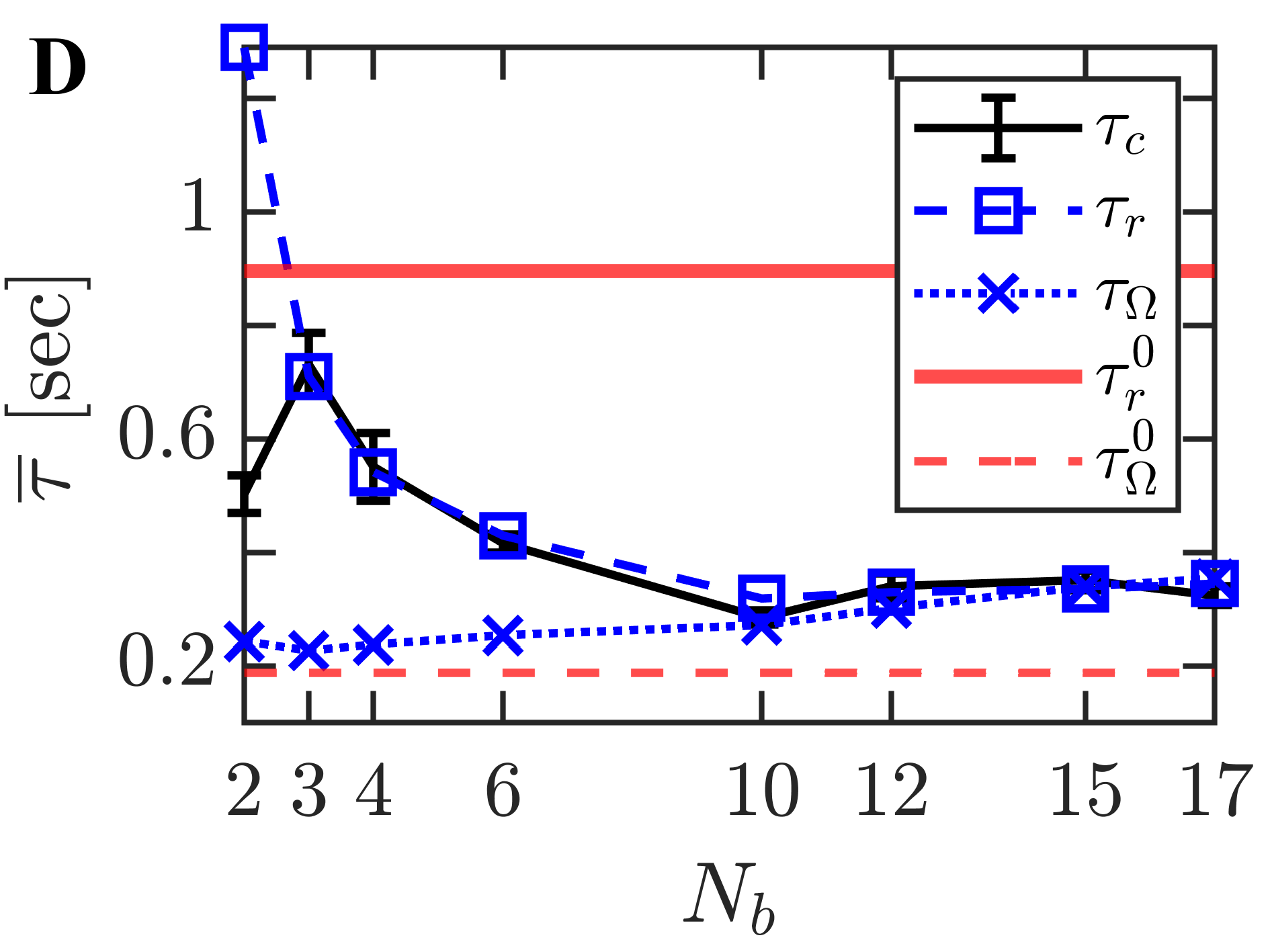}
    \caption{
    \textbf{Stationary dynamics and statistics.}
    We consider configurations of $N_b = \{2,3,4,6,10,12,15,17\}$ bbot baths. 
    \textbf{A.} \ Position probability density (symbols) and the normal (Gaussian) distribution for $N_b=15$ (dashed line).
    \textbf{B.} \ Effective temperature $T_{\text{eff}}$ vs. $N_b$ (upper panel), and corresponding non-Gaussian parameter (lower panel). Error bars are standard deviations.
    \textbf{C.} \ The (ensemble) averaged normalized position autocorrelation $\overline{C_x}(\tau)$ for different $N_b$. 
    \textbf{D.} \ Different characteristic timescales as a function of $N_b$: mean free time between collisions $\tau_c$ (line), steady-state dynamical timescales (obtained by a fit to Eq.~\ref{eq:PCF}) $\tau_r$ (squares) and $\tau_\Omega$ (crosses), the tracer's damping timescale in the absence of bbots $\tau^0_r=0.9$s and $\tau^0_\Omega = \sqrt{m/k}=0.18$s (horizontal lines).
    The results were obtained for an ensemble average of $M=375$ sequences, with the same tracer $m\approx 1$~g, and gravitational stiffness $k\approx 28.2~\text{g/s}^{-2}$.
    }
    \label{fig:3}
\end{figure}

\textit{Range of validity of the FDR.--}
First, we obtain the nonequilibrium steady state dynamics and statistics of the system (Fig. \ref{fig:1}) with different numbers of bbots $N_b$, the results are displayed in Fig. \ref{fig:3}.
We find that $P(x,N_b)$ transitions into a non-Gaussian distribution as $N_b$ decreases (Fig.~\ref{fig:3}\textbf{A}). This is expected as the typical time between collisions $\tau_c$ increases beyond the typical relaxation of the tracer to the bottom of the well \cite{park2020}. 
By using a consistent definition of effective temperature for all configurations, $k_BT_\text{eff} = k\langle \Delta x^2\rangle_0$, we observe a monotonic increase in $T_\text{eff}$ with $N_b$ (Fig. \ref{fig:3}\textbf{B}, upper panel).
We quantify the departure from Boltzmann statistics by calculating the non-Gaussian parameter (excess kurtosis) of the position distribution, $\Gamma=\frac{\langle x^4\rangle}{3\langle x^2\rangle^2}-1$ (Fig. \ref{fig:3}\textbf{B}, lower panel).

In Fig. \ref{fig:3}\textbf{C,D} we plot the normalized autocorrelation $\overline{C_x}(\tau)$ and its characteristic times, $\tau_r$ and $\tau_\Omega$, from a fit to Eq. \ref{eq:PCF}. 
A transition from under- to critically damped dynamics as $N_b$ increases is manifested in the reduction of the oscillations in $\overline{C_x}(\tau)$ and in the relaxation time to the steady state $\tau_r$.  The diminished influence of the tracer's inertia for larger $N_b$ is clearly due to an increase of dissipation caused by more frequent collisions of tracer particle and bbots. The evaluated mean free time between collisions $\tau_c$ results in values similar to $\tau_r$, except for the limiting case of $N_b=2$. 
In the latter, the bbots rarely collide and mainly drive the ball around the trap.
The dissipative timescale associated with the arena was probed by applying the FDR step-perturbation protocol with the bbots removed (see SI$^{\dag}$).
The \textit{passive} damping time $\tau^0_r=0.9$~s was obtained by fitting the non-fluctuating response to Eq. \ref{eq:PCF}, with $\tau^0_\Omega = \sqrt{m/k} = 0.18$~s directly calculated (Fig. \ref{fig:3}\textbf{D}, horizontal lines).
Note that $\tau_\Omega$ shows a non-trivial dependence on $N_b$, due to the physical presence of the bbots in the harmonic trap.

\begin{figure}[t] 
    \centering
    \includegraphics[width=0.3450\textwidth]{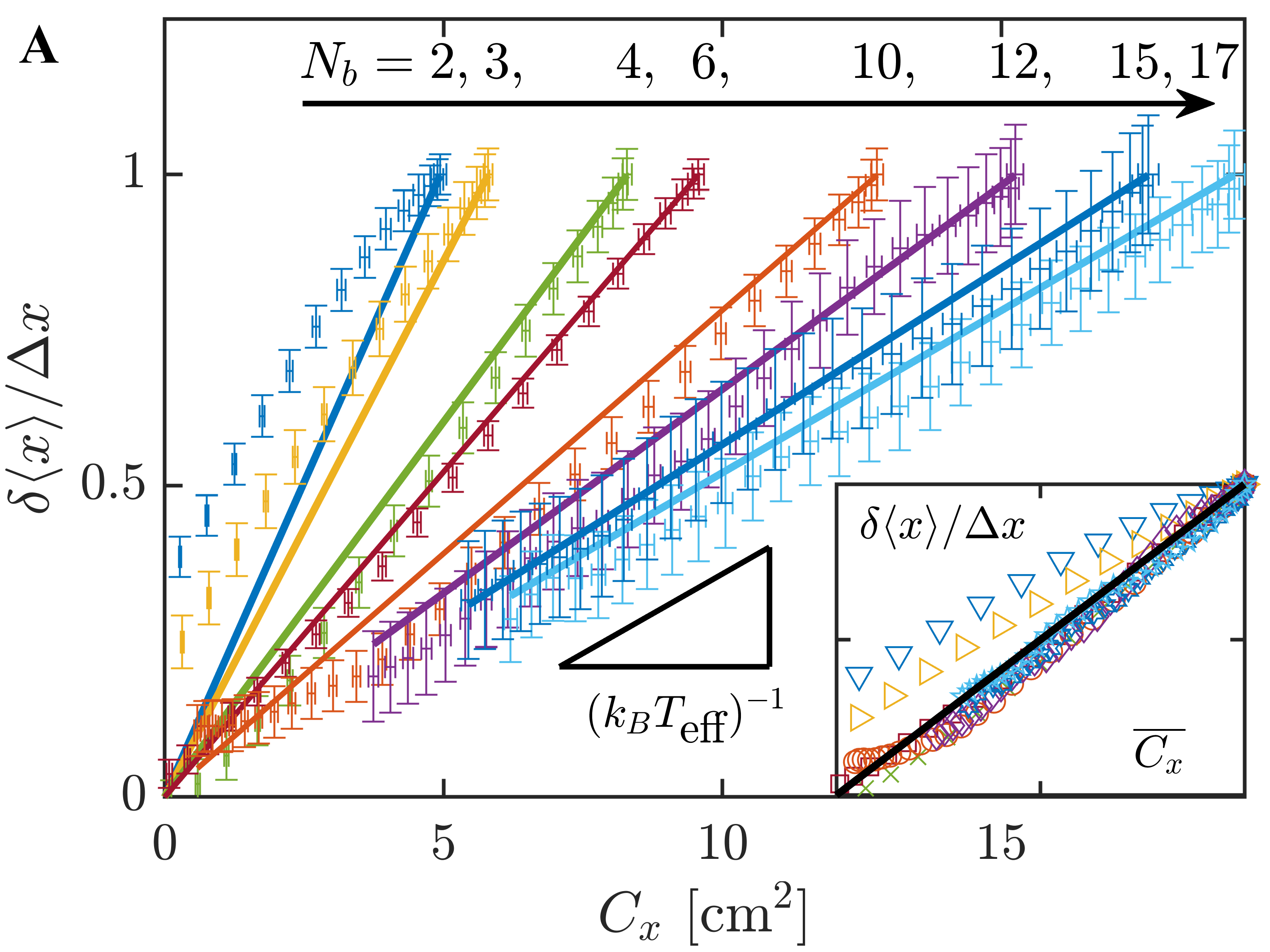}
    
    \includegraphics[width=0.3450\textwidth]{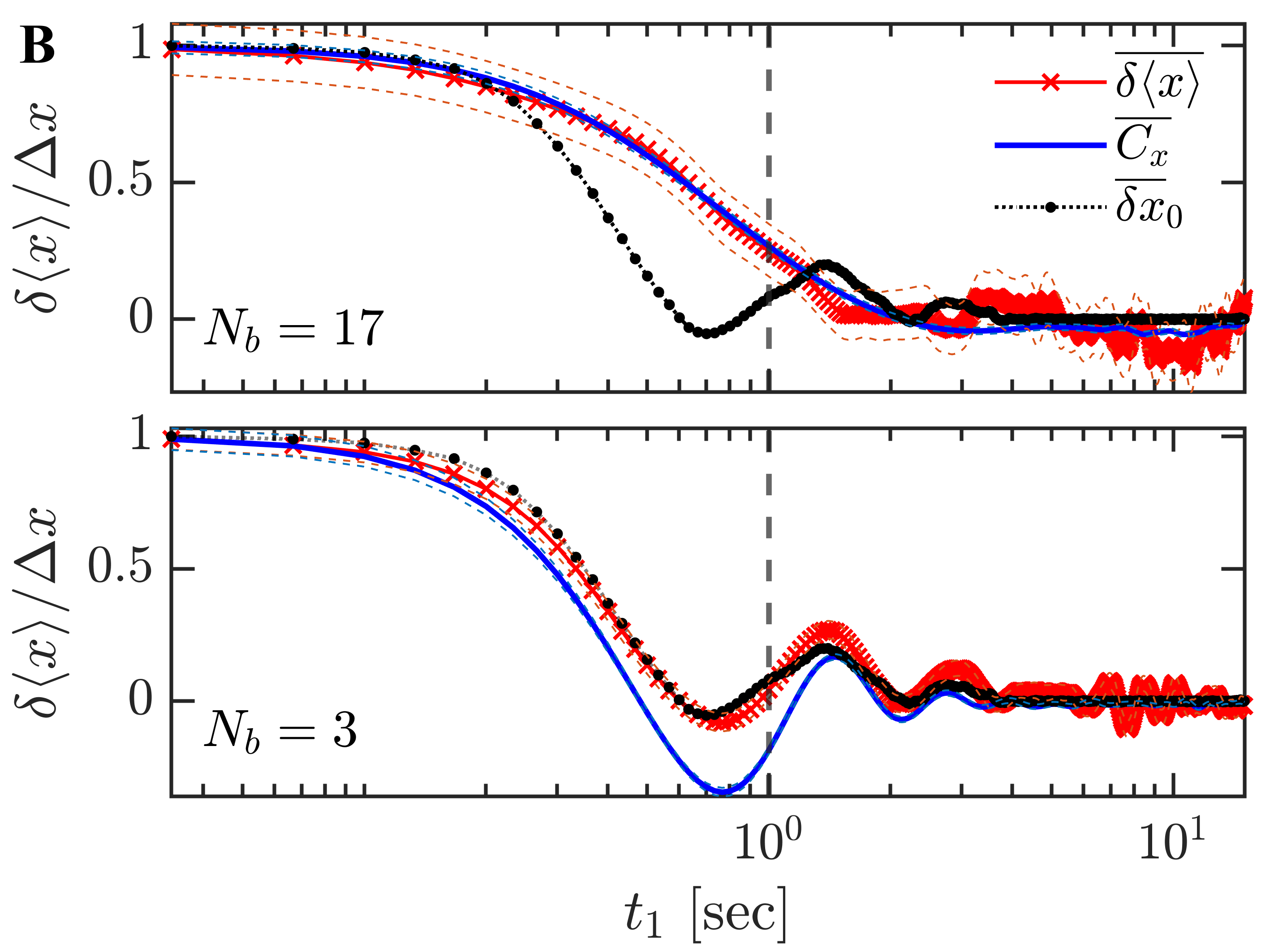}
    \caption{
    \textbf{Validity range of the FDR.}
    The results are for experimental configurations with different $N_b$, as in Fig. \ref{fig:3}.
    \textbf{A.} \ The generalized FDRs are displayed in a parametric plot with the normalized response $\delta\langle x\rangle/\Delta x$ versus the autocorrelation $C_x$, for times $t_1<T_c$.
    Error bars are standard errors.
    The straight linear lines correspond to a perfect FDR with slope $=(k_BT_{\text{eff}})^{-1}$.
    The inset shows a unitary parametric plot, where a deviation (FDR violation) is observed for lowest bbot numbers $N_b=3$ and $2$ (triangles). 
    \textbf{B.} \ Normalized plot of generalized FDRs for $N_b=17$ (upper panel) and $N_b=3$ (lower panel). Here, $\delta x(t_1)/\Delta x$ is plotted against $\overline{C_x}(t_1)$, and is compared to the normalized non-fluctuating response $\overline{\delta x}_0$ under the same perturbation ($10$V fan voltage), in the absence of bbots.
    See SI$^{\dag}$ for all dynamic FDR results.
    }
    \label{fig:4}
\end{figure}

Fig. \ref{fig:4}\textbf{A} shows a parametric plot of the generalized FDRs for different $N_b$. The normalized mean response $\delta \langle x\rangle/\Delta x$ is plotted versus the autocorrelation $C_x$, within the response time regime $t_1<T_c$. Rescaling the autocorrelation by $T_{\text{eff}}$, we obtain a master curve (Fig.~ \ref{fig:4}\textbf{A}, inset) in which all the systems obeying the FDR collapse to a single line. A clear deviation is seen for $N_b=2$ and $3$, which violates the FDR.
In our experiments, it is clear that the main source of the tracer's fluctuations is the collisions with the bbots. To determine the dominant dissipation source we compare the (normalized) mean response $\delta \langle x \rangle$ of the two extreme cases, $N_b=17$ and $N_b=3$, to the deterministic response of the tracer in the absence of bbots $\delta x_0$ (Fig. \ref{fig:4}\textbf{B}, upper and lower panel, respectively). 
For $N_b=17$ collisions are sufficiently frequent ($\tau_c<\tau^0_r$) such that the mean response deviates significantly from $\delta x_0$. 
In this regime, the dissipation is predominantly governed by the collisions between the tracer and the bbots. 
In contrast, for $N_b=3$ collisions are rare ($\tau_c\approx \tau^0_r$), and the mean response of the tracer closely resembles $\delta x_0$. While collisions play a role in the dissipation, the primary source of dissipation in this case is the friction between the tracer and the bowl. Consequently, this scenario leads to a violation of the FDR.

\textit{Discussion.--}
In summary, we have experimentally confirmed the validity of a generalized FDR, far from equilibrium.
We find that for $N_b>3$ the FDR holds with an effective temperature that is given by the tracer's mean potential energy, $T_{\text{eff}}\sim \langle \Delta x^2\rangle_0$, even when the steady state dynamics are underdamped and the probability distribution is non-Gaussian. 
We note that at $N_b>17$, bbots frequently tumble and form long-lived static clusters at the center trap region. This is reflected in a small flat region in the center of the tracer's position distribution for $N_b=17$ (see SI$^{\dag}$). The presence of the bbot cluster changes the trap's properties and the tracer's dynamics,  hindering accurate FDR tests for $N_b>17$.  

Moreover, our light tracer can move out of the plane, evading entrapment within large clusters of bbots.
Therefore, in contrast to previous reports in which caging and memory effects have been shown to result in FDR violations at high bath densities \cite{baldovin2022review,gnoli2014}, we did not observe such a violation in our experiments. 
However, we expect to see FDR violations once collisions are not sufficiently frequent to act as the main source for dissipation to external perturbations. This is the case for the lowest numbers of bbots $N_b=2$ and $3$, where the generalized FDR fails. 

Our findings for a single tracer system suggest that the FDR should hold for many tracer systems if tracer-tracer interactions are non-dissipative. However, previous evidence of FDR violations in dense active assemblies \cite{Levis2015} and reported non-trivial effects of bath density on tracer-bath interactions \cite{caprini2023} point to a more complex situation.

The confirmation of a generalized FDR, across an extensive range of effective temperatures, raises the following three broader issues.  
Is there an overarching equipartition relation that connects these effective temperatures to $T_\text{eff}$ defined via momentum variables rather than positional ones? 
Preliminary results indicate that a linear relation between $T_\text{eff}$ and the tracer's mean kinetic energy, $\frac{1}{2} m\langle v^2\rangle_0$, holds up to $N_b=10$ in our experiments (see SI$^{\dag}$), in accord with a previous study \cite{chastaing2017}.
Secondly, do these temperatures hold significance within the thermodynamic framework governing these systems? For instance, could they dictate energy flow in scenarios where two such systems are in contact?
Lastly, one feature of our experimental system that was observed in a previous study is that its dynamics are consistent with Markovian dynamics \cite{engbring2023}.
While Markovianity is not essential for the FDR to prevail in thermal equilibrium, an intriguing avenue of exploration involves assessing its validity in weakly perturbed systems immersed in an active, non-Markovian environment, where dissipation and fluctuation derive from the same underlying physical process.

\section*{Conflicts of interest}
There are no conflicts to declare.

\section*{Acknowledgements}
DB and YR acknowledge support from the European Research Council (ERC) under the European Union’s Horizon 2020 research and innovation program (Grant agreement No. 101002392). 



%

\clearpage

\appendix

\section*{Supplementary Information}

\subsection*{The bbot bath characteristics}

We characterize the properties of the bbot \textit{active gas} in the harmonic trap, and its effect on the motion of a passive tracer.
A single bbot in a parabolic arena adopts a circular path when it has sufficient inertia \cite{dauchot2019}.
Additionally, the bbot favors clockwise rotation reflecting its inherent chirality, which stems from an asymmetric motor. 

\begin{figure}[h] 
    \centering
    \includegraphics[width=0.155\textwidth]{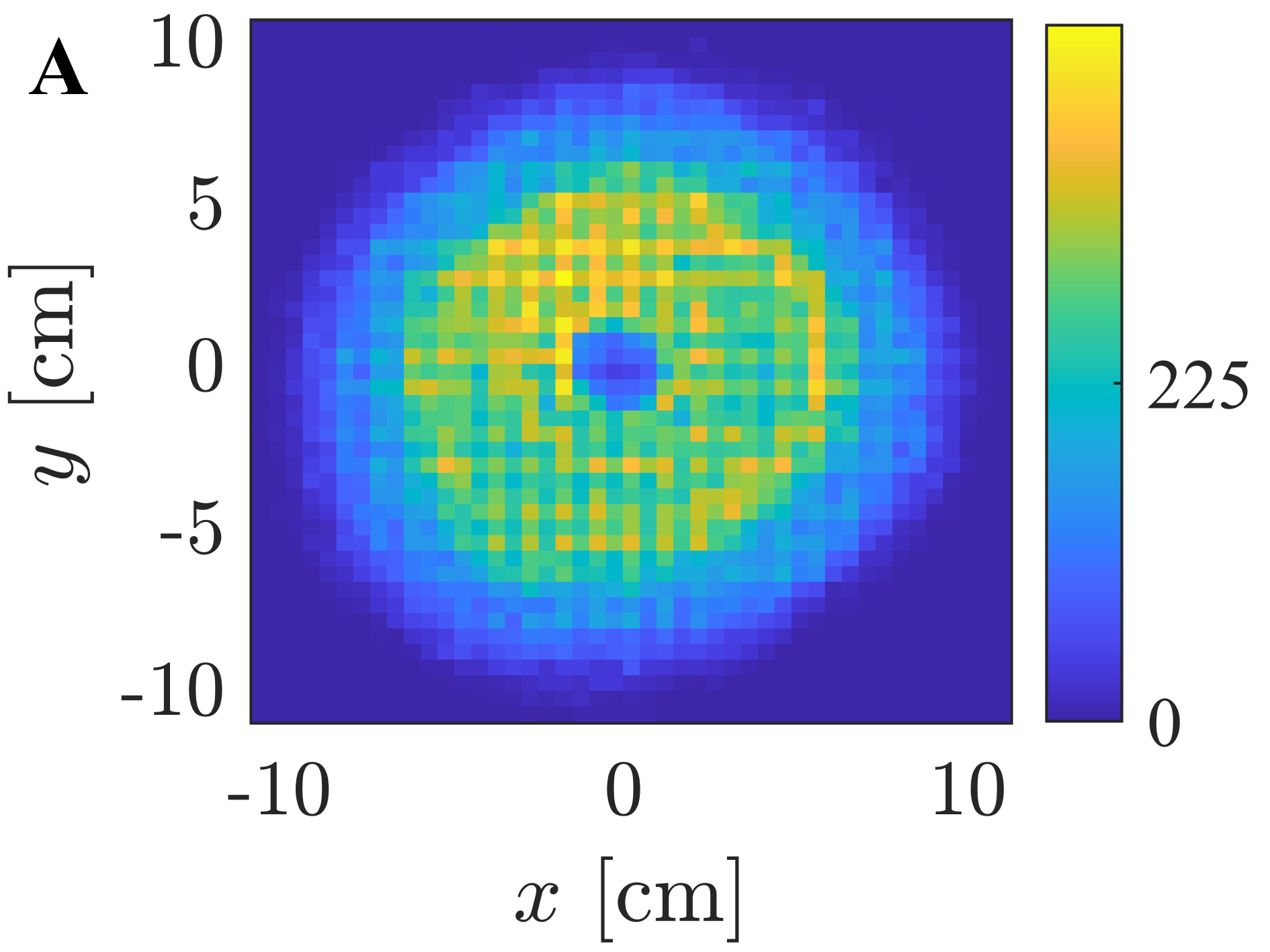}
    \includegraphics[width=0.155\textwidth]{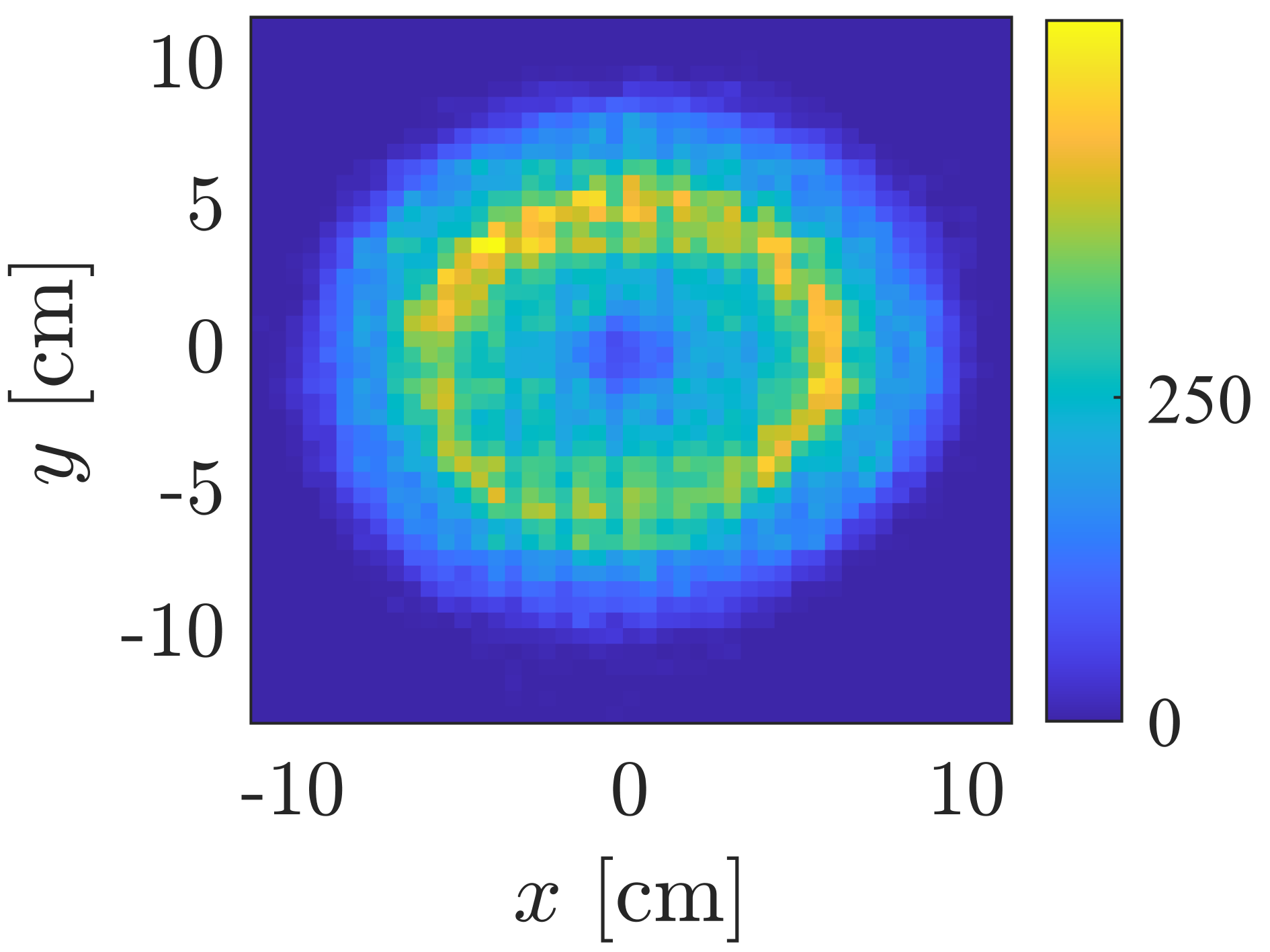}  
    \includegraphics[width=0.155\textwidth]{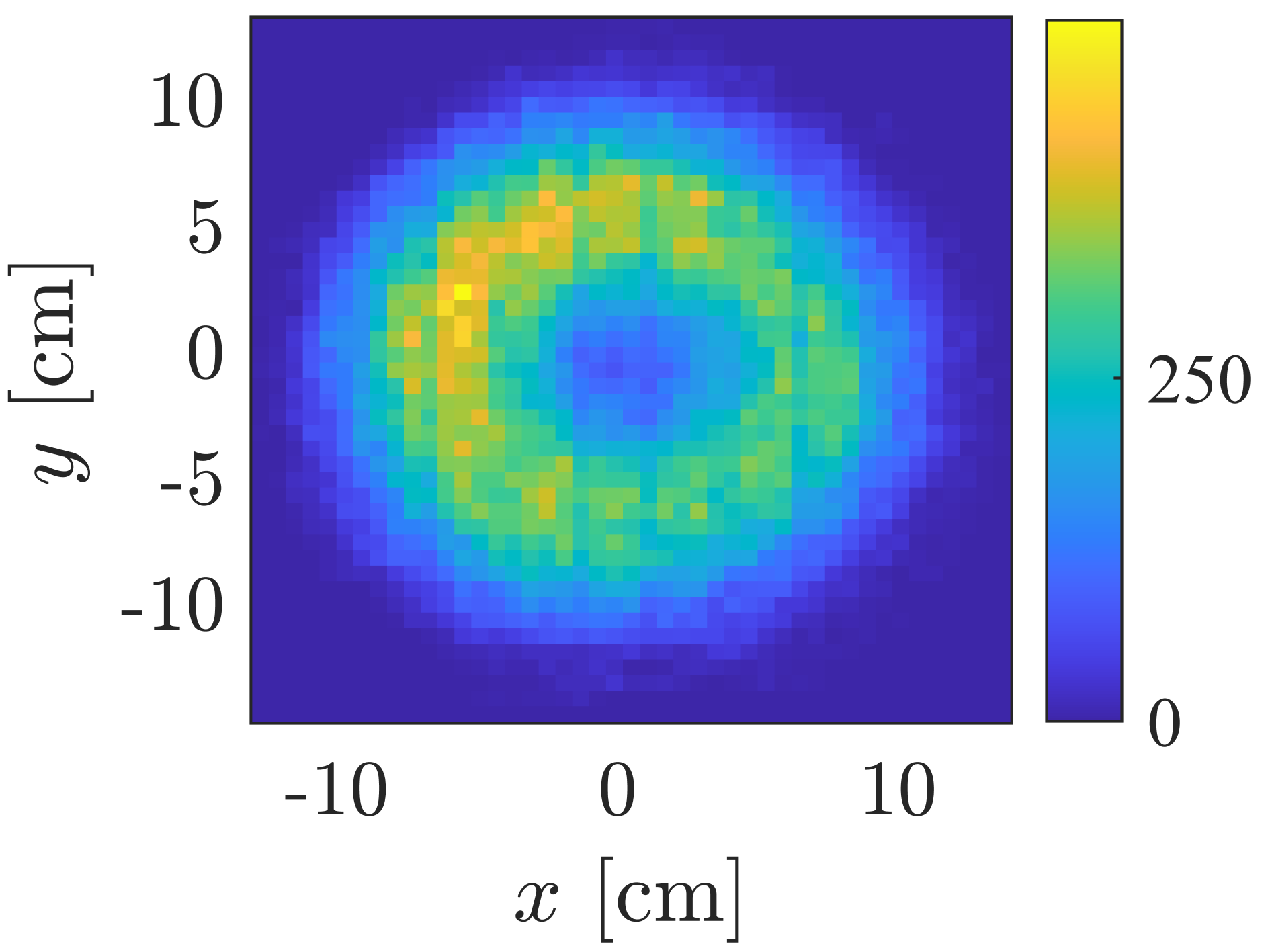}
    \includegraphics[width=0.155\textwidth]{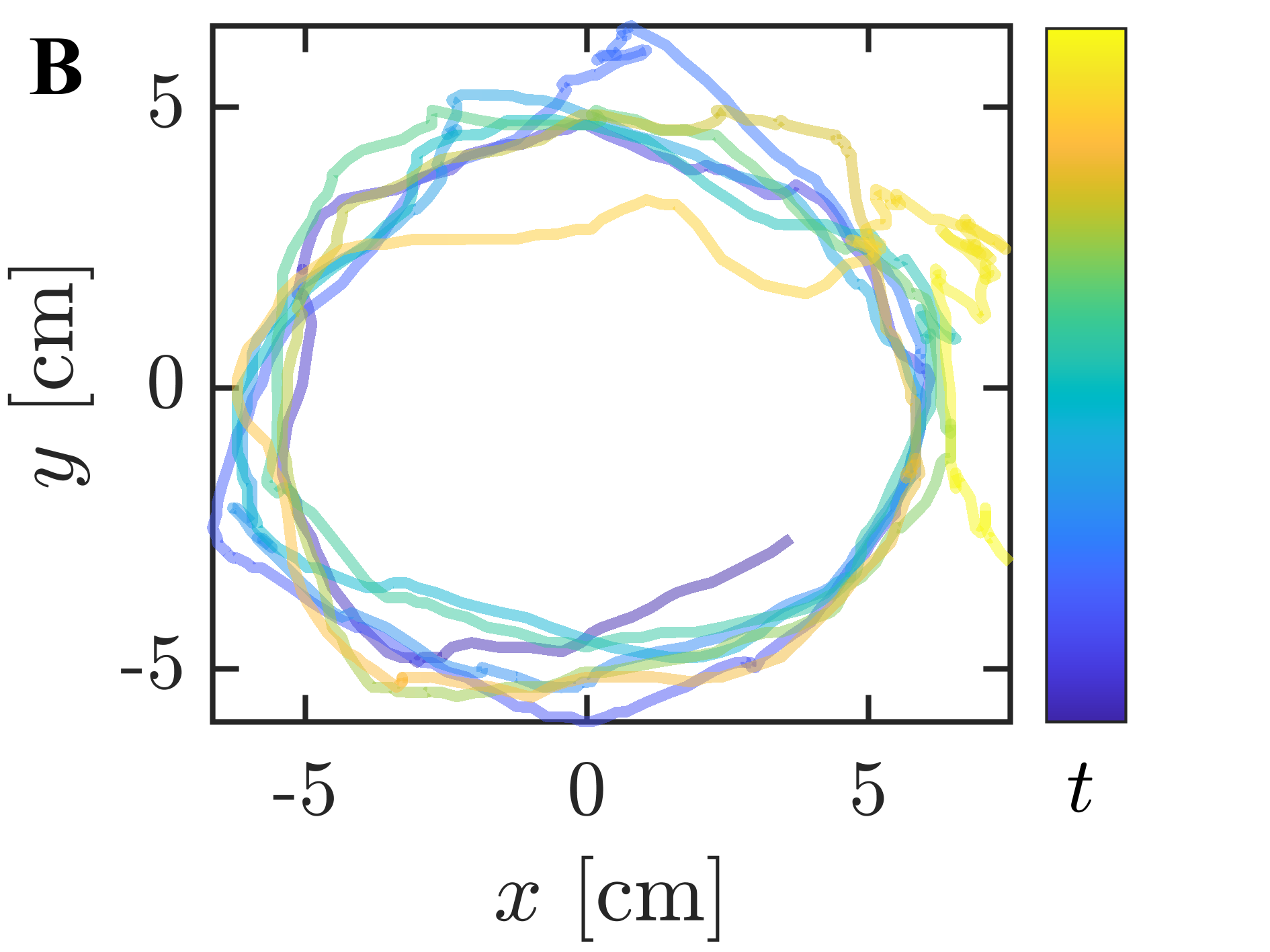}
    \includegraphics[width=0.155\textwidth]{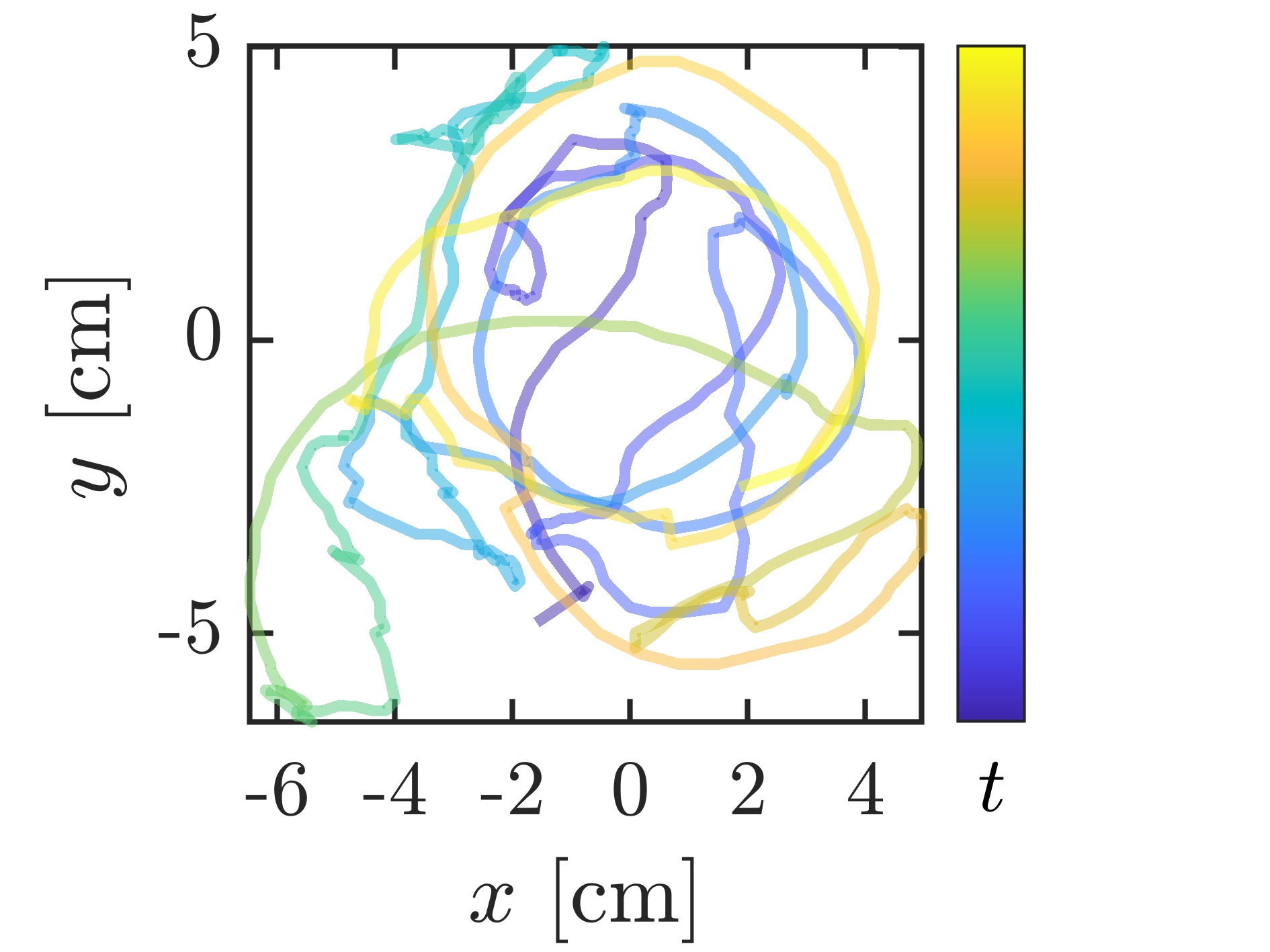}  
    \includegraphics[width=0.155\textwidth]{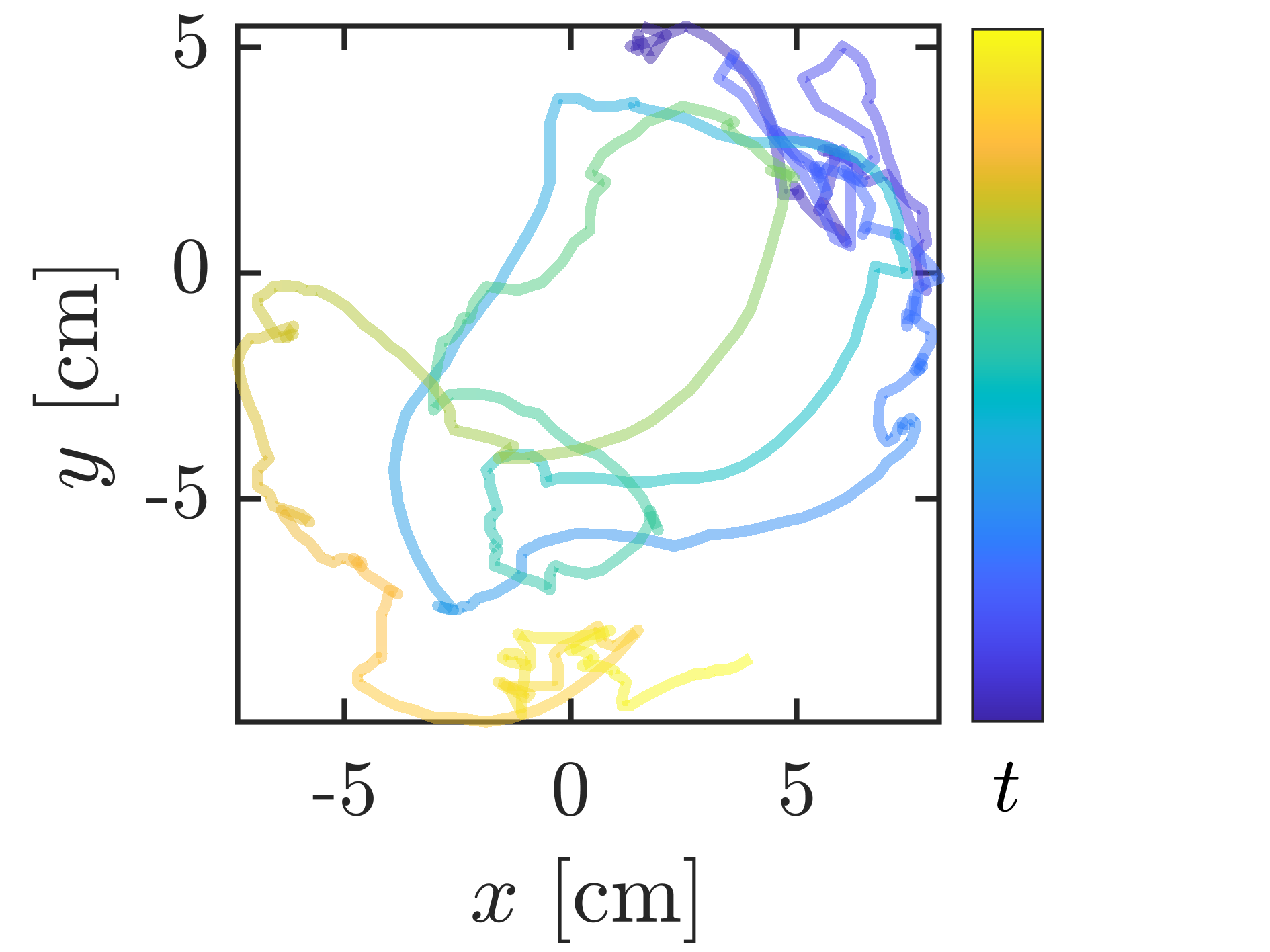} 
    \includegraphics[width=0.18\textwidth]{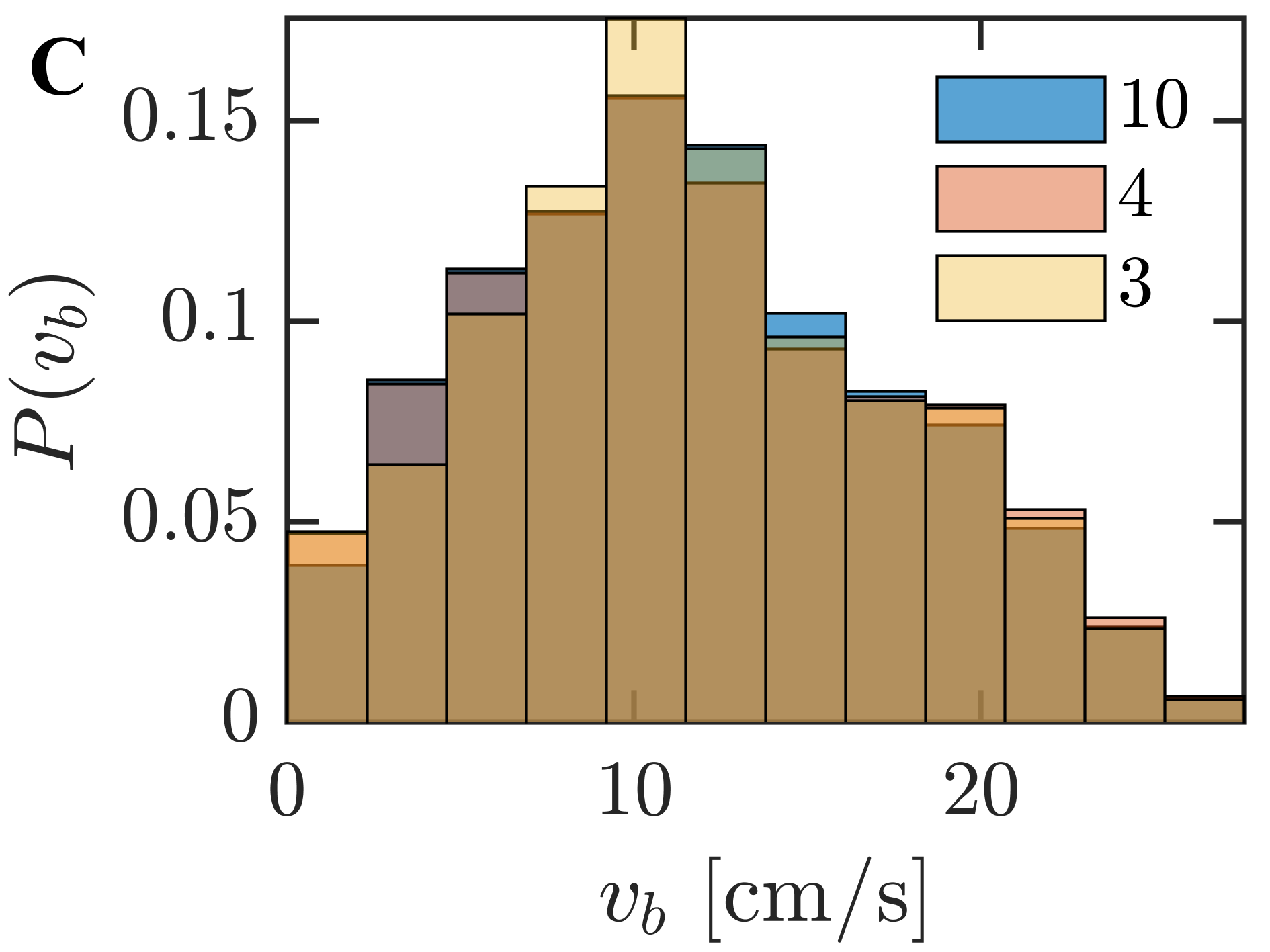}
    \includegraphics[width=0.18\textwidth]{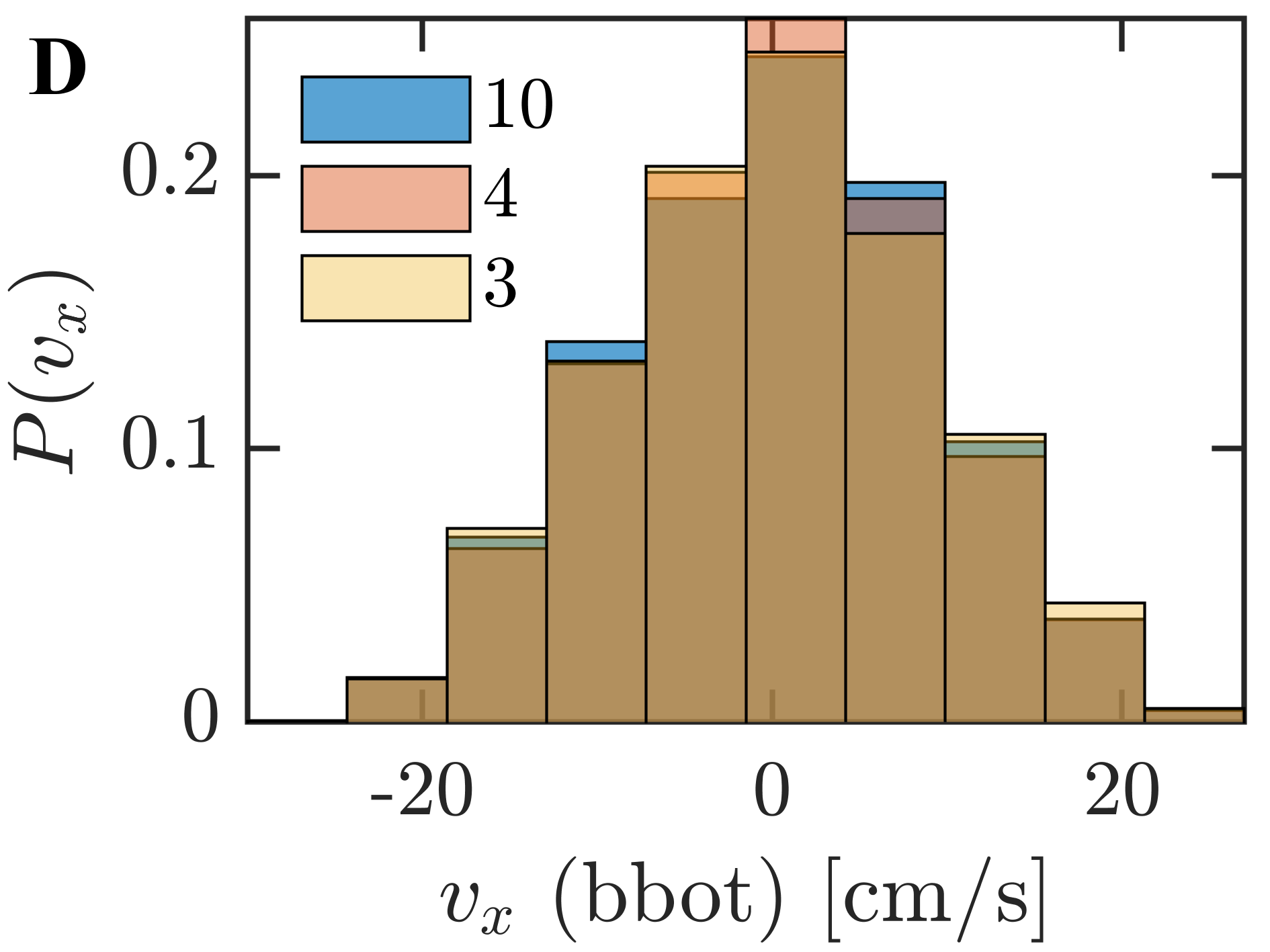}  
    \caption{
    \textbf{Distributions and typical trajectories of bbots for different densities.} 
    \textbf{A.} \ The total ensemble positional distribution of the bbots in the harmonic trap. 
    The results are for $N_b=3$, $4$, and $10$, from left to right respectively,
    obtained by tracking all of the bbots for 30 minutes.
    \textbf{B.} \ Typical single bbot trajectories, $t\in[0,15]s$.
    The bbot speed (\textbf{C.}) and velocity (\textbf{D.}) distributions remain independent of $N_b$. }
    \label{fig:SIfig1} 
\end{figure}

\begin{figure}[H] 
    \centering
    \includegraphics[width=0.155\textwidth]{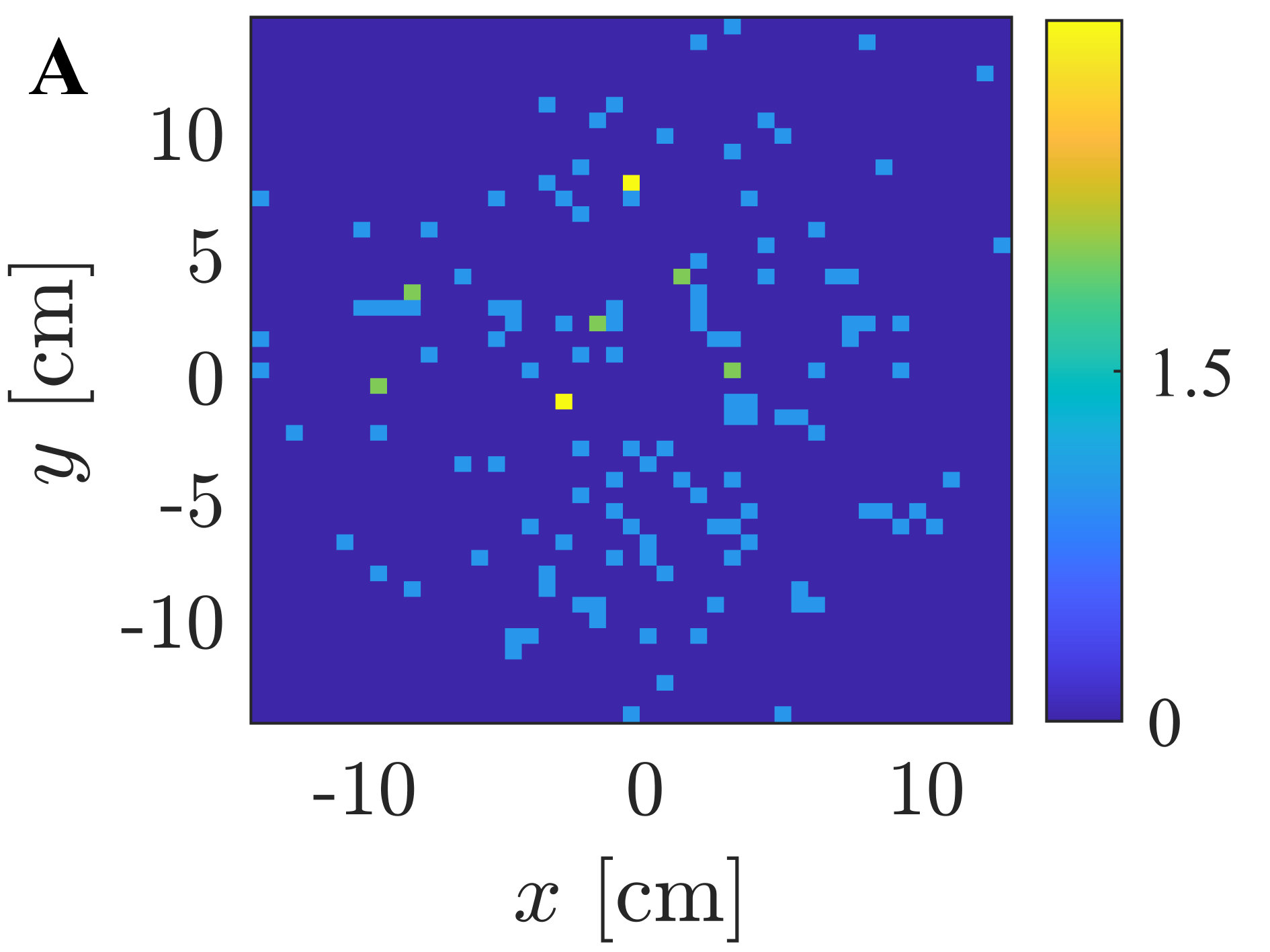}
    \includegraphics[width=0.155\textwidth]{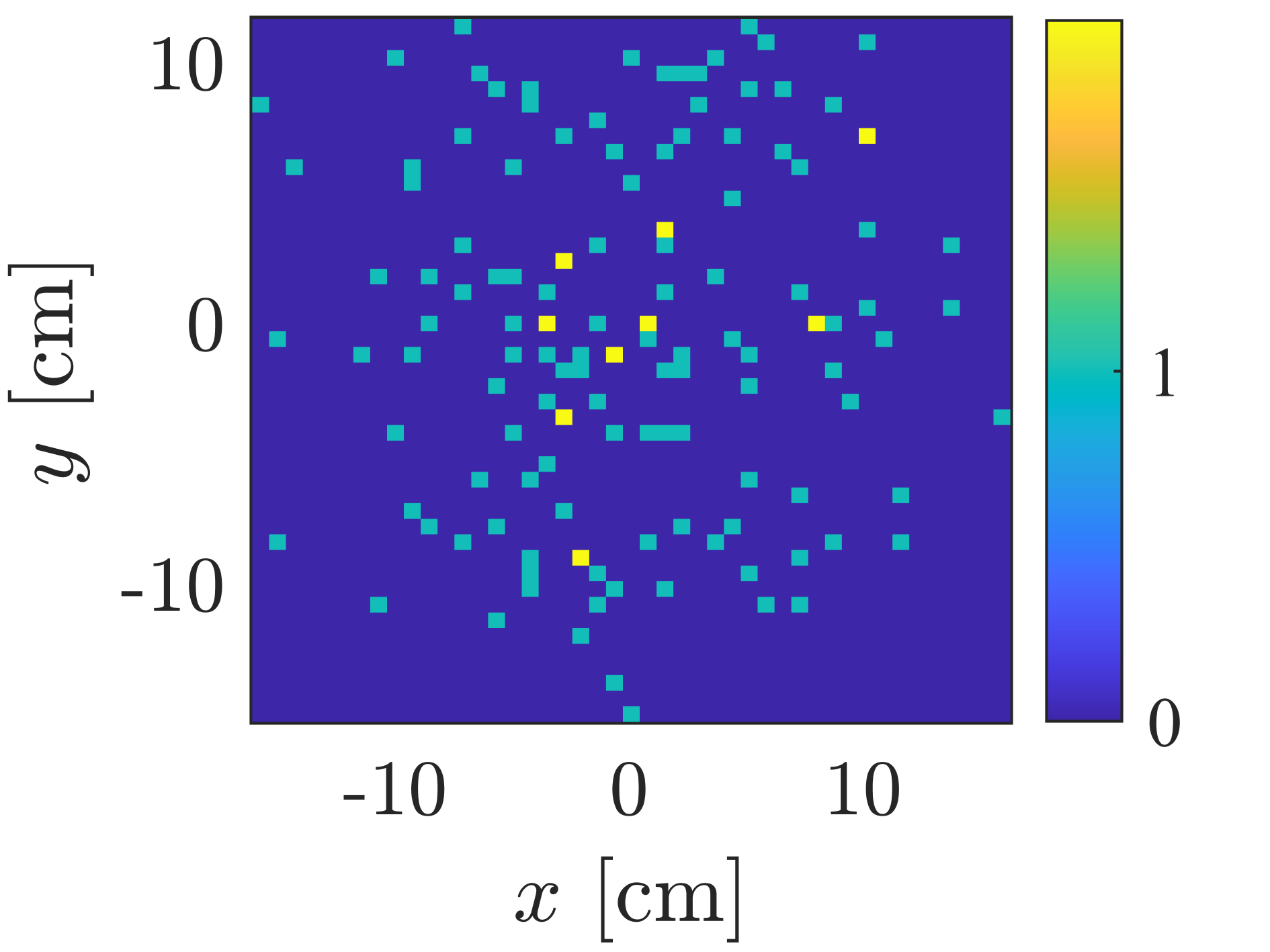}
    \includegraphics[width=0.155\textwidth]{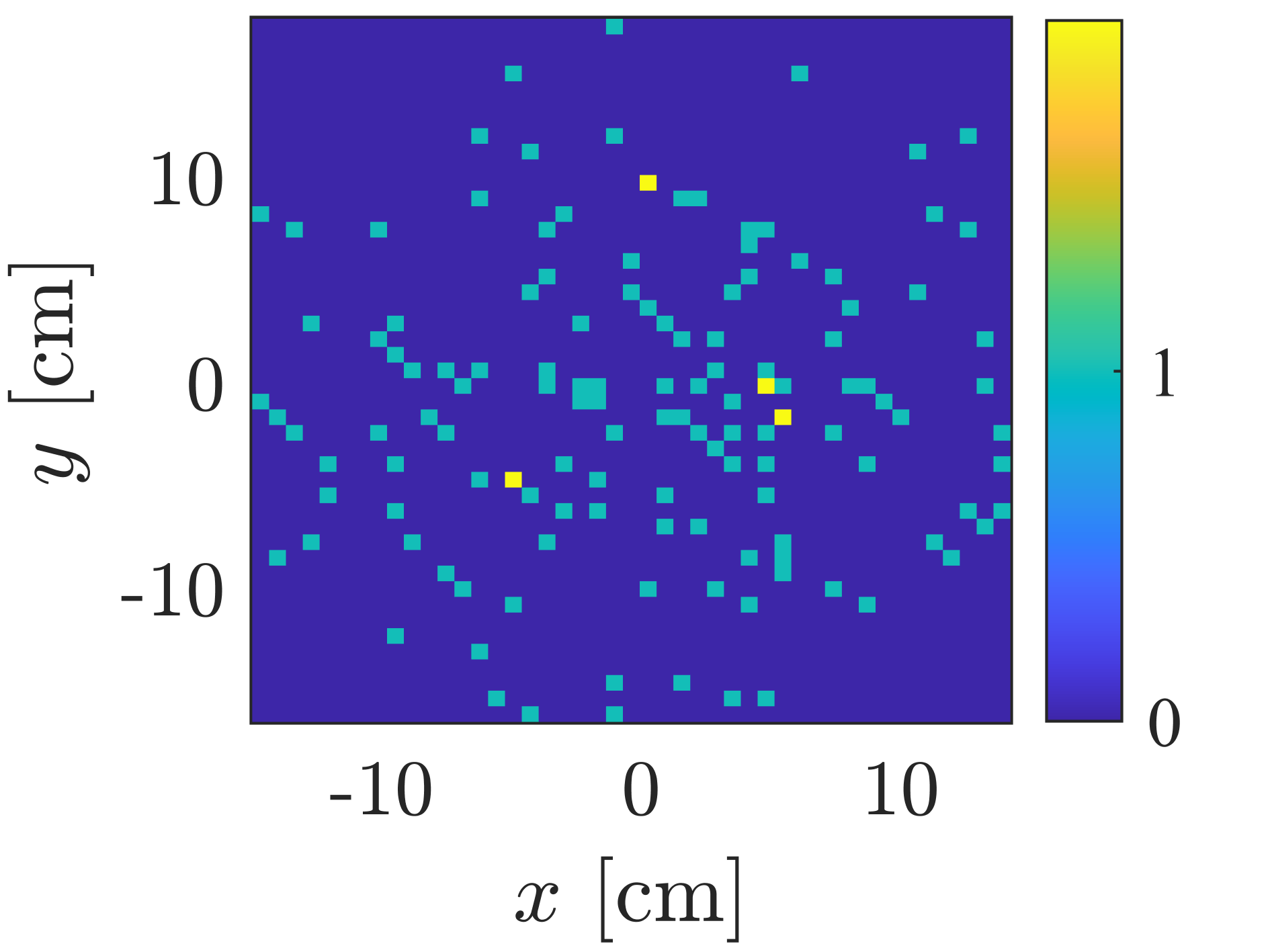}
    \caption{
    \textbf{The external air-stream does not affect the bbots.} 
    We show the difference of the spatial bbot distributions with the fan on and off, for $N_b=3$, $4$, and $10$, from left to right, respectively, showing that the bbots are not affected by the fan's air-flow.  }
    \label{fig:SIfig2}
\end{figure}

By introducing additional bbots into the harmonic trap, collisions between them randomize their motion. This effect increases with the bbot number $N_b$ (see Fig.~\ref{fig:SIfig1}\textbf{A} and \textbf{B}).
We observe that the speed and velocity distributions remain independent of $N_b$, with a mean speed $\langle v_b\rangle \approx 11$cm/s (see Fig.~\ref{fig:SIfig1}\textbf{C} and \textbf{D}).
We also find that the bbots are heavy enough to ensure no air-stream effect on their motion during a response measurement (see Fig.~\ref{fig:SIfig2}). 

\begin{figure}[H] 
    \centering
    \includegraphics[width=0.18\textwidth]{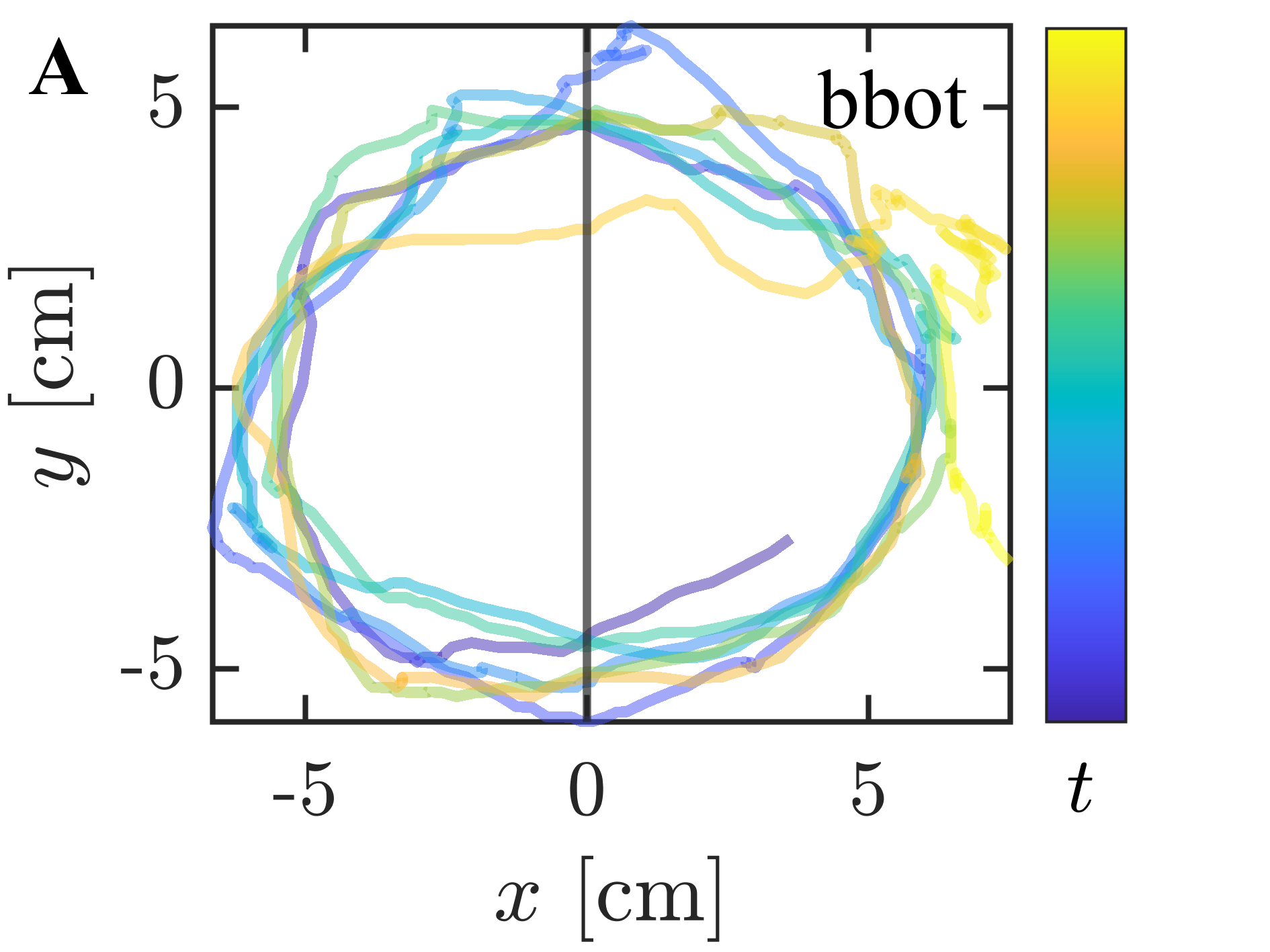}
    \includegraphics[width=0.18\textwidth]{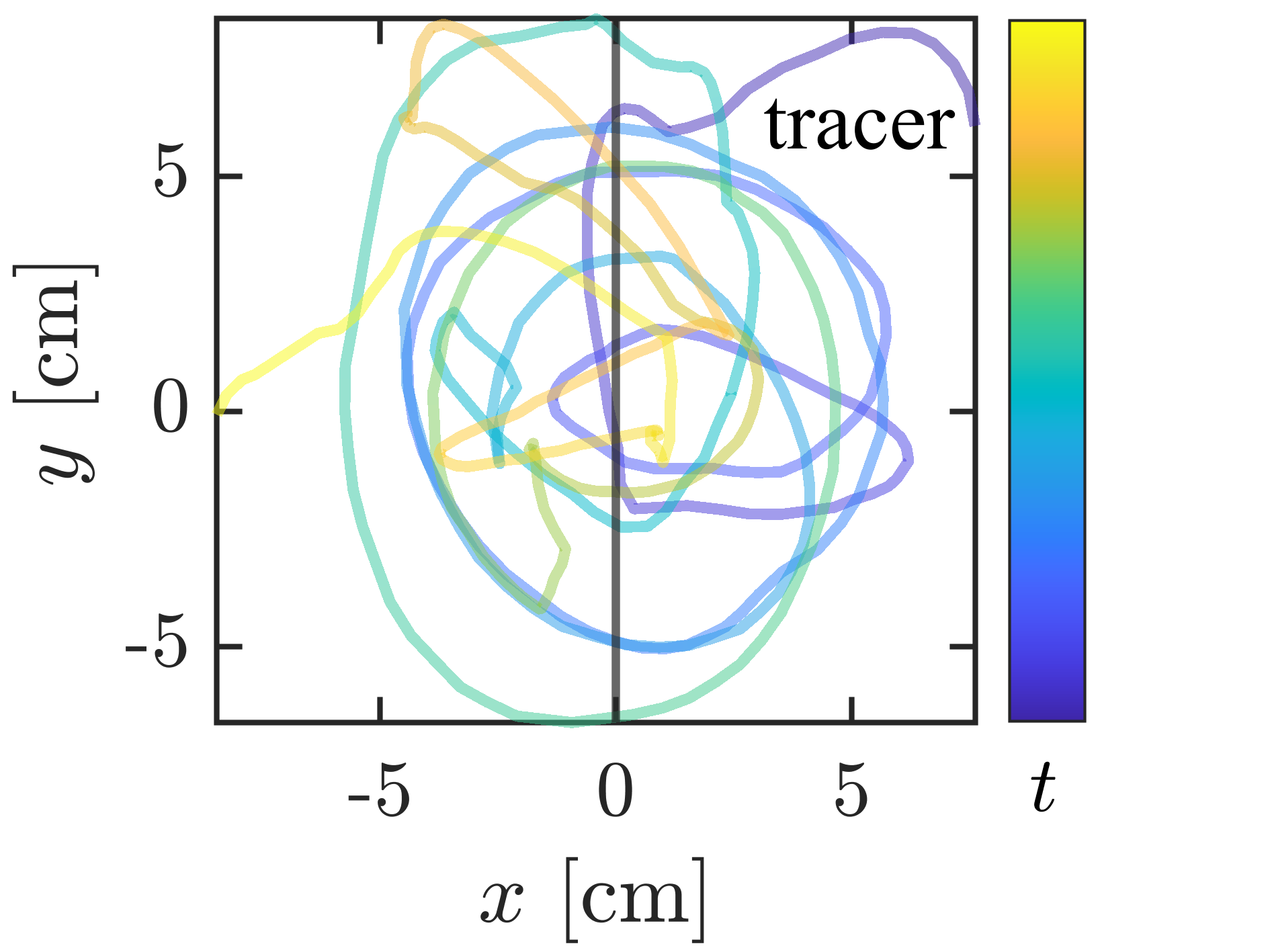}
    \includegraphics[width=0.18\textwidth]{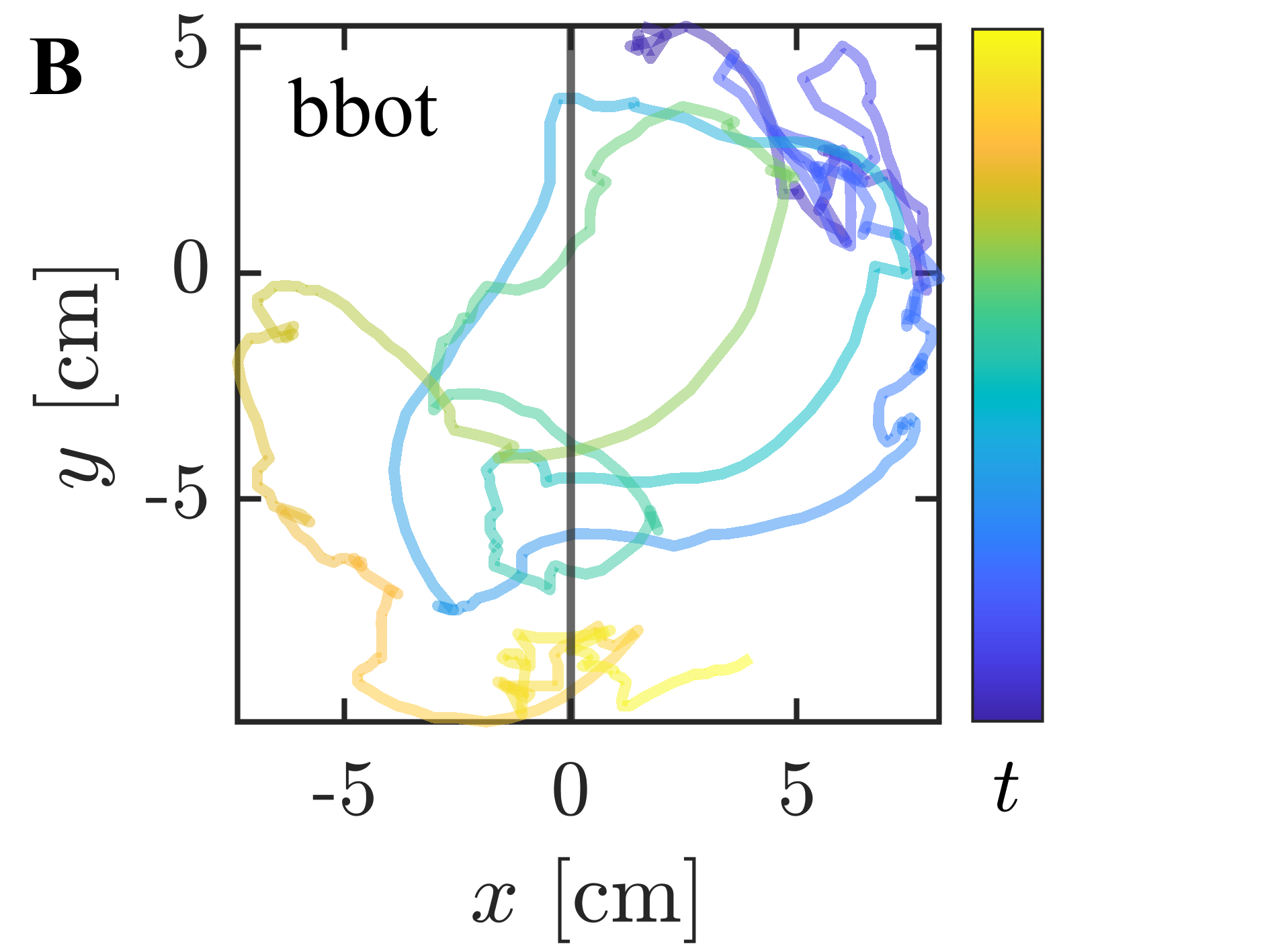} 
    \includegraphics[width=0.18\textwidth]{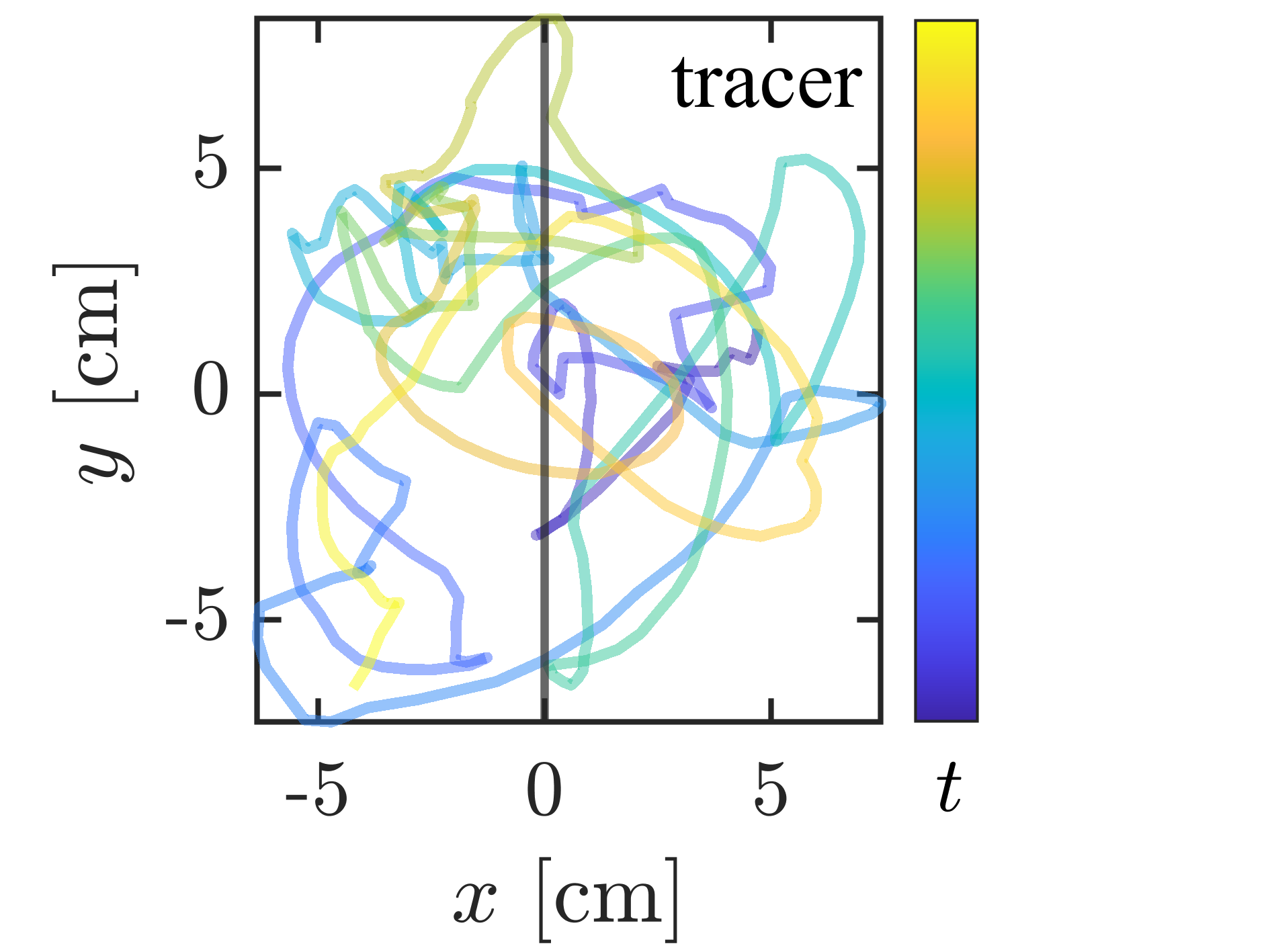}
    \includegraphics[width=0.18\textwidth]{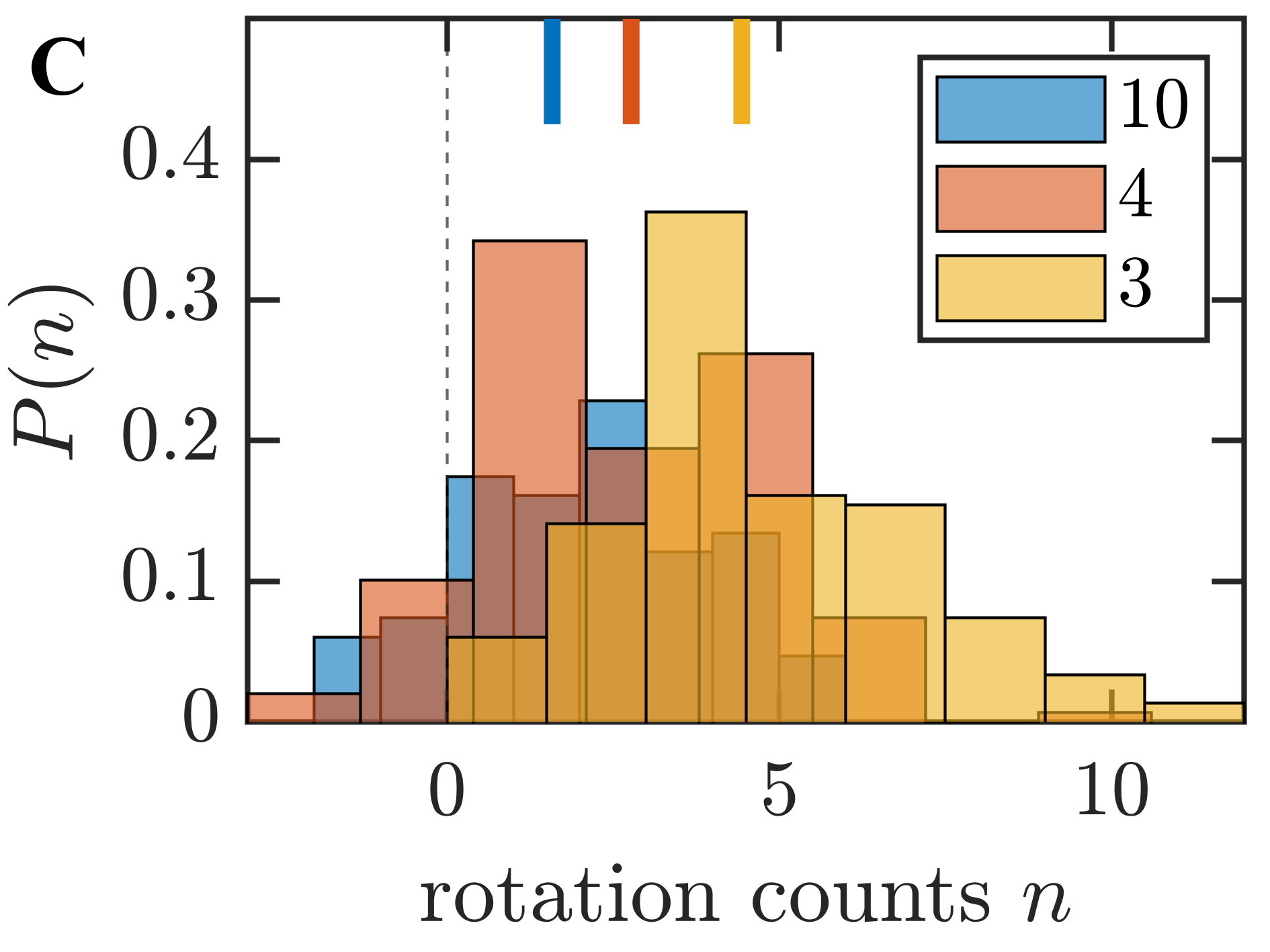}
    \includegraphics[width=0.18\textwidth]{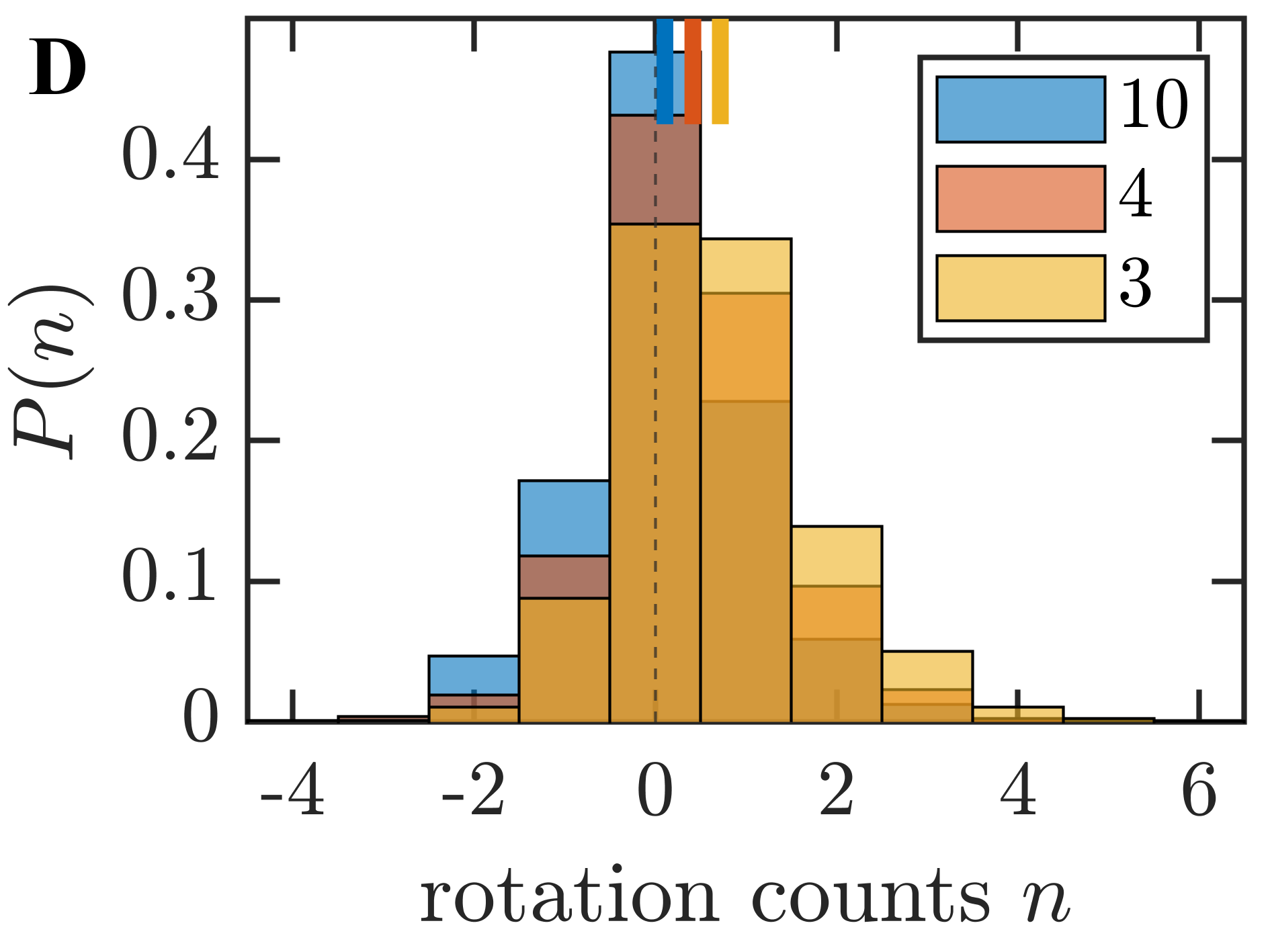}  
    \caption{
    \textbf{The system's chirality.} 
    Typical single bbot and tracer trajectories, in a system with $N_b=3$ (\textbf{A}) and $N_b=10$ (\textbf{B}).
    \textbf{C.} \ The distribution of bbot instantaneous rotations $\pm n$ across a central line at $x=0$, where $n>0$ indicates favored clockwise direction.
    Obtained by tracking 200 bbots trajectories ($t\in[0,5]s$).
    The upper lines are the means of $P(n)$, showing that net chirality is reduced by increasing $N_b$.
    \textbf{D.} \ The same chirality analysis for the passive tracer. }
    \label{fig:SIfig3}
\end{figure}

In Fig.~\ref{fig:SIfig3}\textbf{A}, \textbf{B} we compare typical trajectories of a bbot and a tracer under the same setup.
Note that the tracer's motion is randomized even for $N_b=3$, where collisions are rare.
We evaluate the net chirality of the bbot bath by counting the clockwise ($+1$) and anticlockwise ($-1$) crossings of a central line at $x=0$.
By tracking 200 bbot trajectories ($t\in[0,5]$s), 
we obtain the sum of bbot rotations $\pm n$ per frame, and a probability distribution $P(n)$ (see Fig.~\ref{fig:SIfig3}\textbf{C}).
We find that the systems attain a net clockwise chirality that is reduced by increasing $N_b$.
In comparison, we provide the chirality of the tracer particle's motion. The latter results in small values and vanishes for high $N_b$ (Fig.~\ref{fig:SIfig3}\textbf{D}). 

\subsection*{A passive tracer in a bbot bath}

\begin{figure}[H] 
    \centering
    \includegraphics[width=0.20\textwidth]{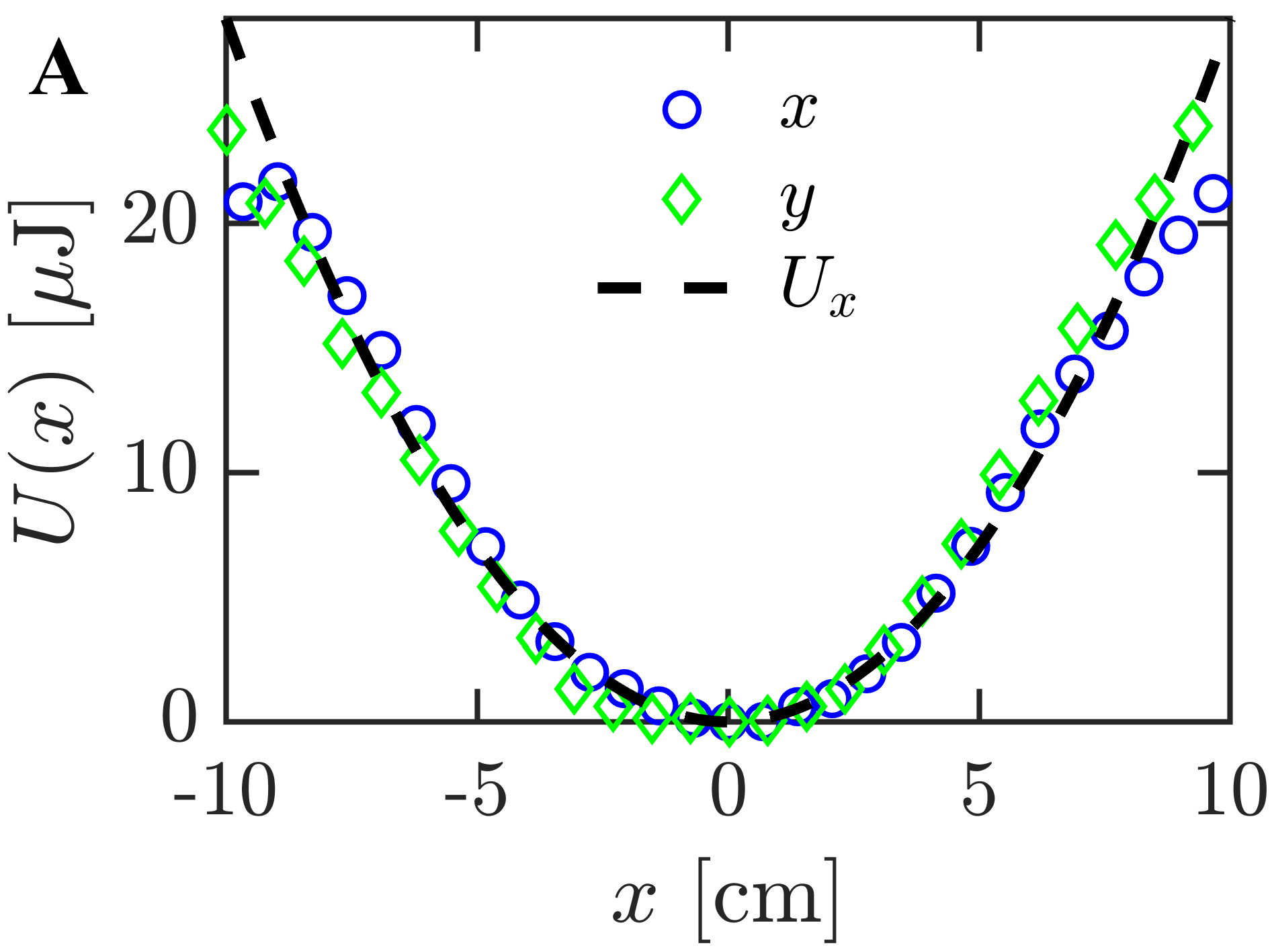}
    \includegraphics[width=0.20\textwidth]{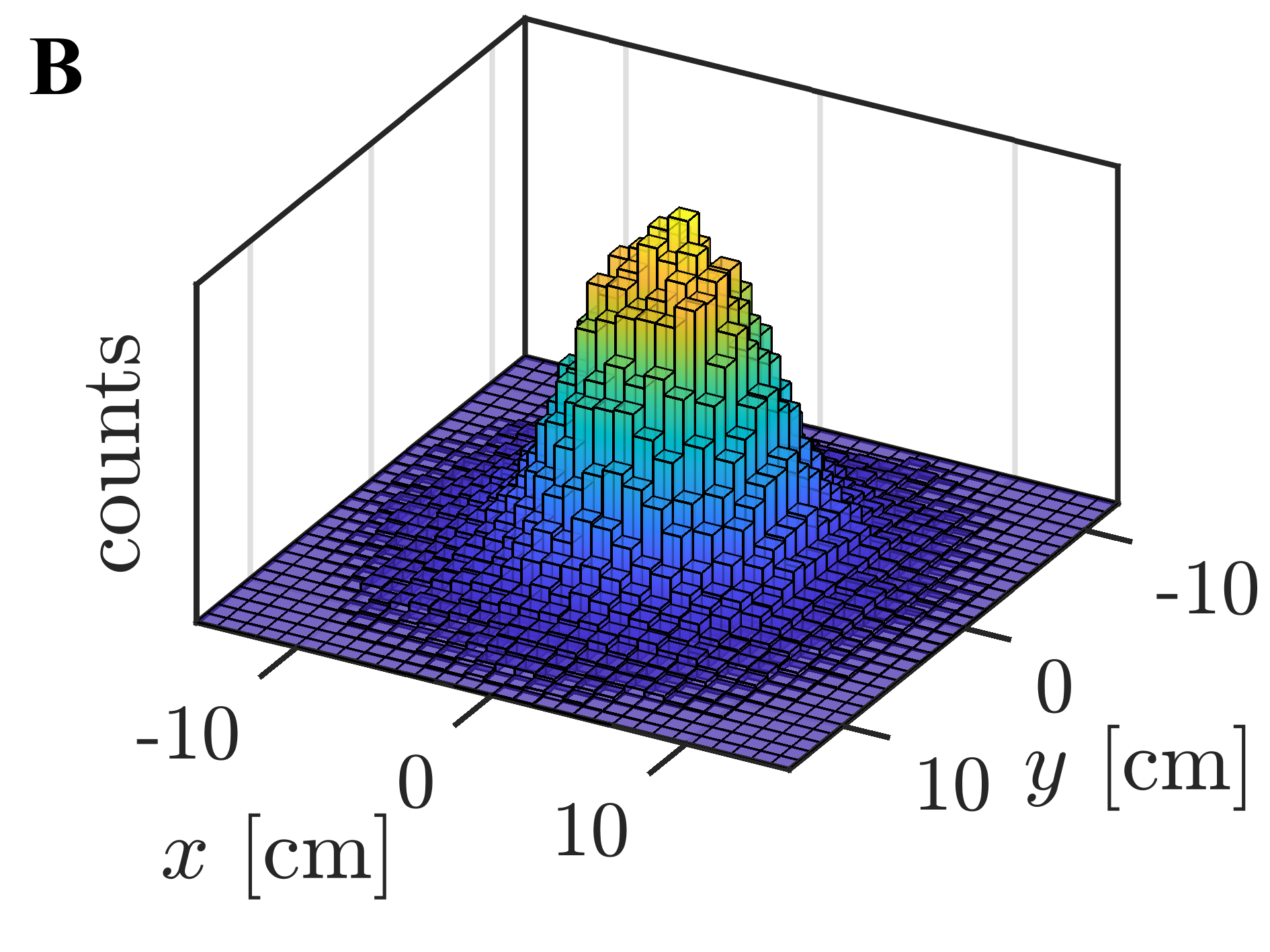} 
    \caption{
    \textbf{Spatial probability distribution and effective potential.} 
    \textbf{A.} \ The physical trapping potential $U_x(x)=mgax^2$ (dashed line), compared to the Boltzmann statistics based experimental estimate of the potential. The results are obtained by averaging $M=375$ tracer position time-sequences, with $t\in[0,30]$~s, in a $N_b=15$ bbot bath.
    \textbf{B.} \ A 3D ($x,y$) tracer position histogram. }
    \label{fig:SIfig4}
\end{figure}

The tracer performs a Brownian-like motion, with a Boltzmann-like PDF for $N_b=15$ (Fig.~\ref{fig:SIfig4}).
The main effect of the bath density on the tracer is an increase in the number of collisions and the energy transferred to it.
Thus, once the collision frequency is sufficient, the PDF reflects the gravitational potential of the bowl. 

\begin{figure}[H] 
    \centering
    \includegraphics[width=0.20\textwidth]{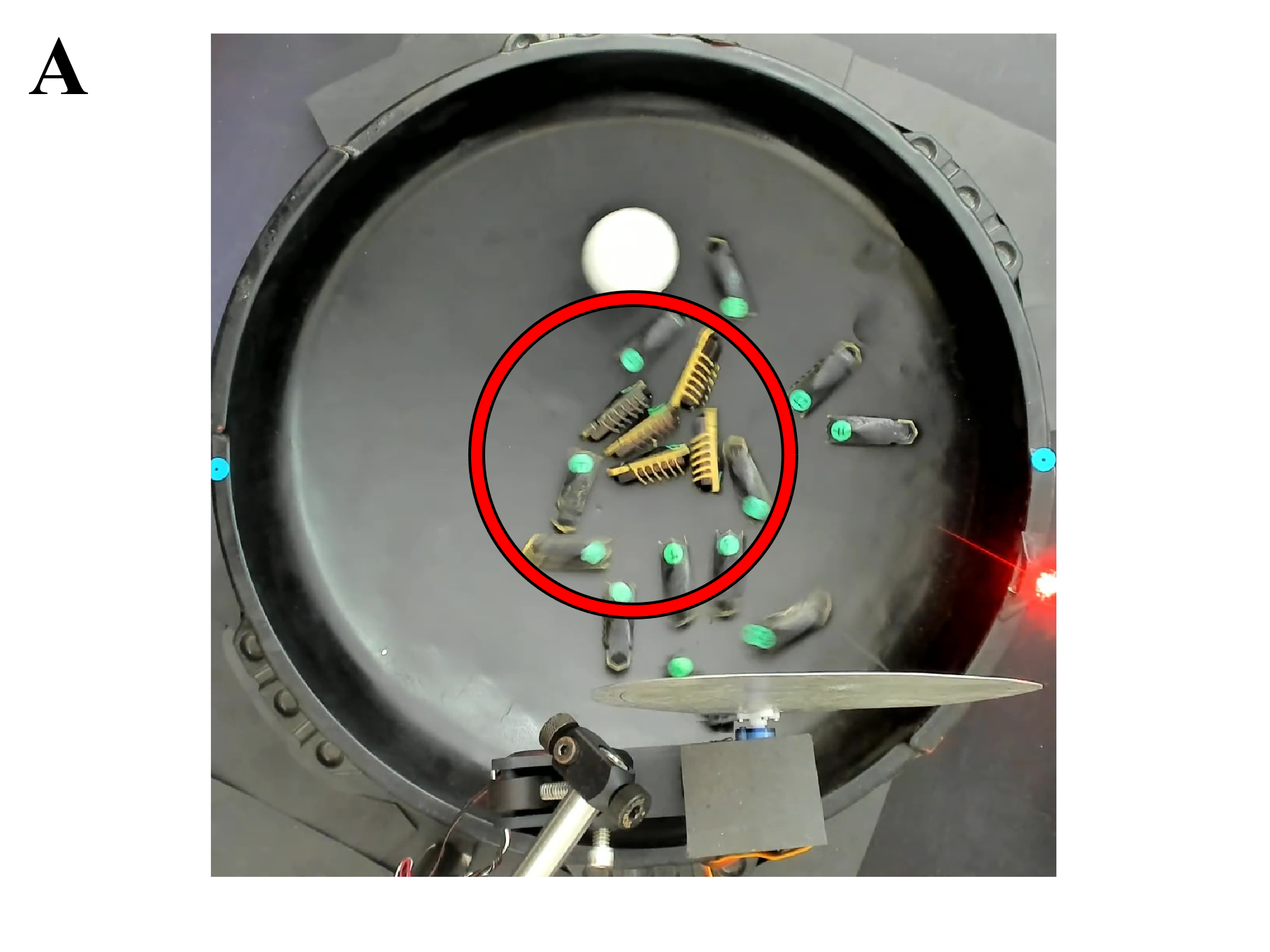}    
    \includegraphics[width=0.20\textwidth]{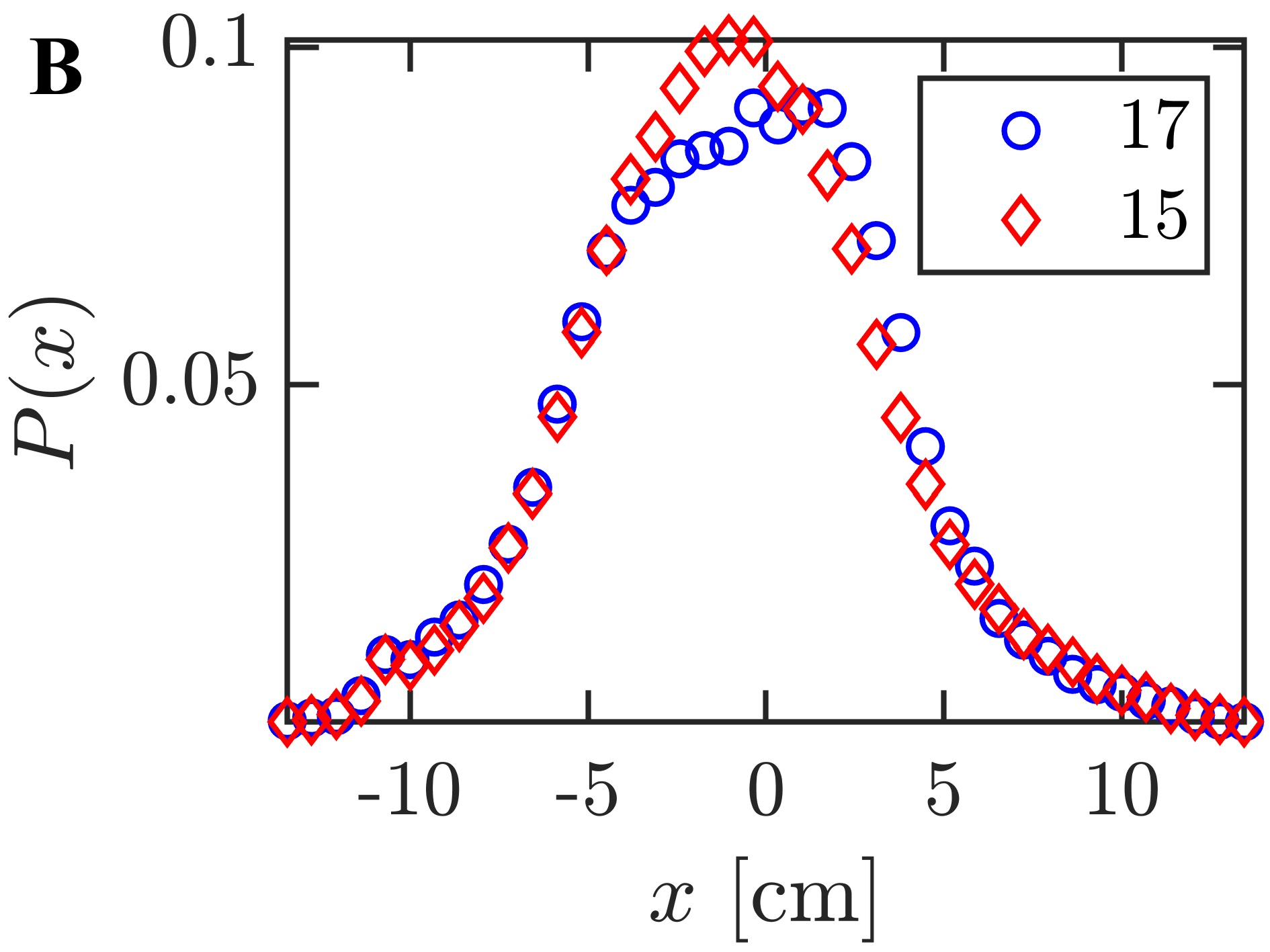}    
    \caption{
    \textbf{A.} \ Snapshot of an experiment with $N_b=17$, showing a temporal static cluster of tumbled bbots in the center trap region.
    \textbf{B.} \ The tracer's position probability distribution for $N_b=17$, showing a flat region in the center, reflecting the physical exclusion by static bbot clusters. }
    \label{fig:SIfig5}
\end{figure}

\begin{figure}[H] 
    \centering
    \includegraphics[width=0.20\textwidth]{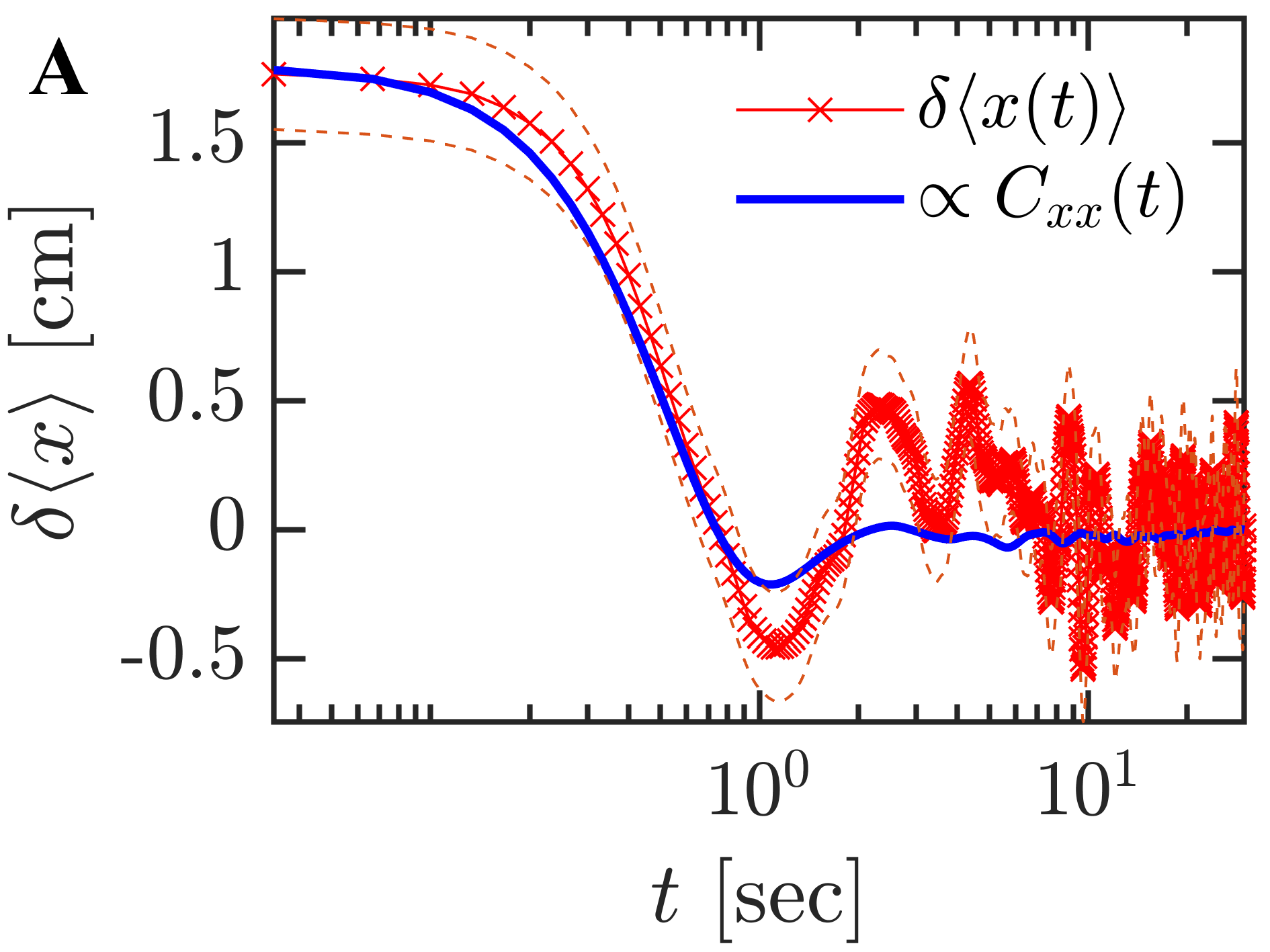}
    \includegraphics[width=0.20\textwidth]{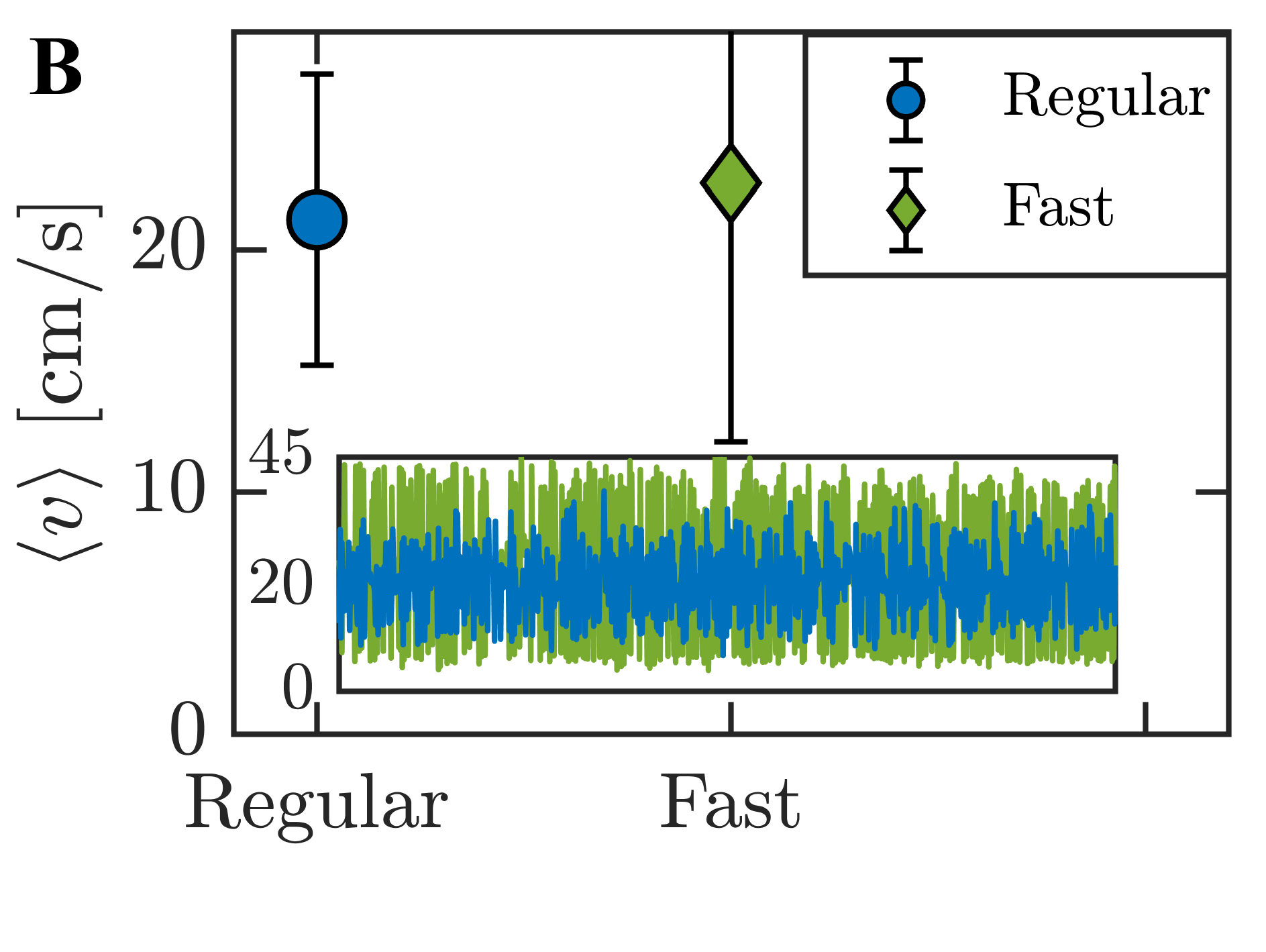}
    \caption{ 
    \textbf{FDR test with faster bbots.} 
    \textbf{A.} \ Generalized FDR test with $k_BT_{\text{eff}}=k\langle x^2\rangle_0$ for $N_b=6$ fast bbots, $9$~V fan operating voltage (weak perturbation), and an average over $M=375$ perturbation sequences. \ 
    \textbf{B.} Average individual speed of a single bbot walking in a parabolic arena. Regular bbot (circle), as used in our work, and a fast bbot (diamond), as used in panel (A). 
    The inset displays typical sequences of instantaneous speeds $v(t)$ in cm/s for regular (blue) and fast (green) bbots.  }
    \label{fig:SIfig11}
\end{figure}

We note that a bbot can both tumble (fall on its side) and get back up, due to random collisions with other bbots.
Once $N_b$ becomes sufficiently large, tumbled bbots form long-lived static clusters at the center trap region (see an example in Fig.~\ref{fig:SIfig5}\textbf{A}).
A similar phenomenon is observed in parabolic arenas with different sizes and curvatures.
The latter can prevent the tracer from accessing the center trap region.
In turn, for $N_b=17$, we observe a flat region in the center of the tracer's position distribution (Fig.~\ref{fig:SIfig5}\textbf{B}).
Nevertheless, the FDR holds in this case.
Due to this technical issue, we cannot conduct a meaningful FDR test above $N_b>17$ and, therefore, cannot exclude FDR violations for higher densities.
Alternatively, we observe that the FDR holds by changing the bbot speed but keeping $N_b$ constant. 
Namely, the FDR is valid with $T_{\text{eff}}\sim \langle \Delta x\rangle_0$ for $N_b=6$ \textit{fast bbots} (see Fig.~\ref{fig:SIfig11}), suggesting the validity of the FDR for varying bath particle activities. 

\begin{figure}[H] 
    \centering
    \includegraphics[width=0.20\textwidth]{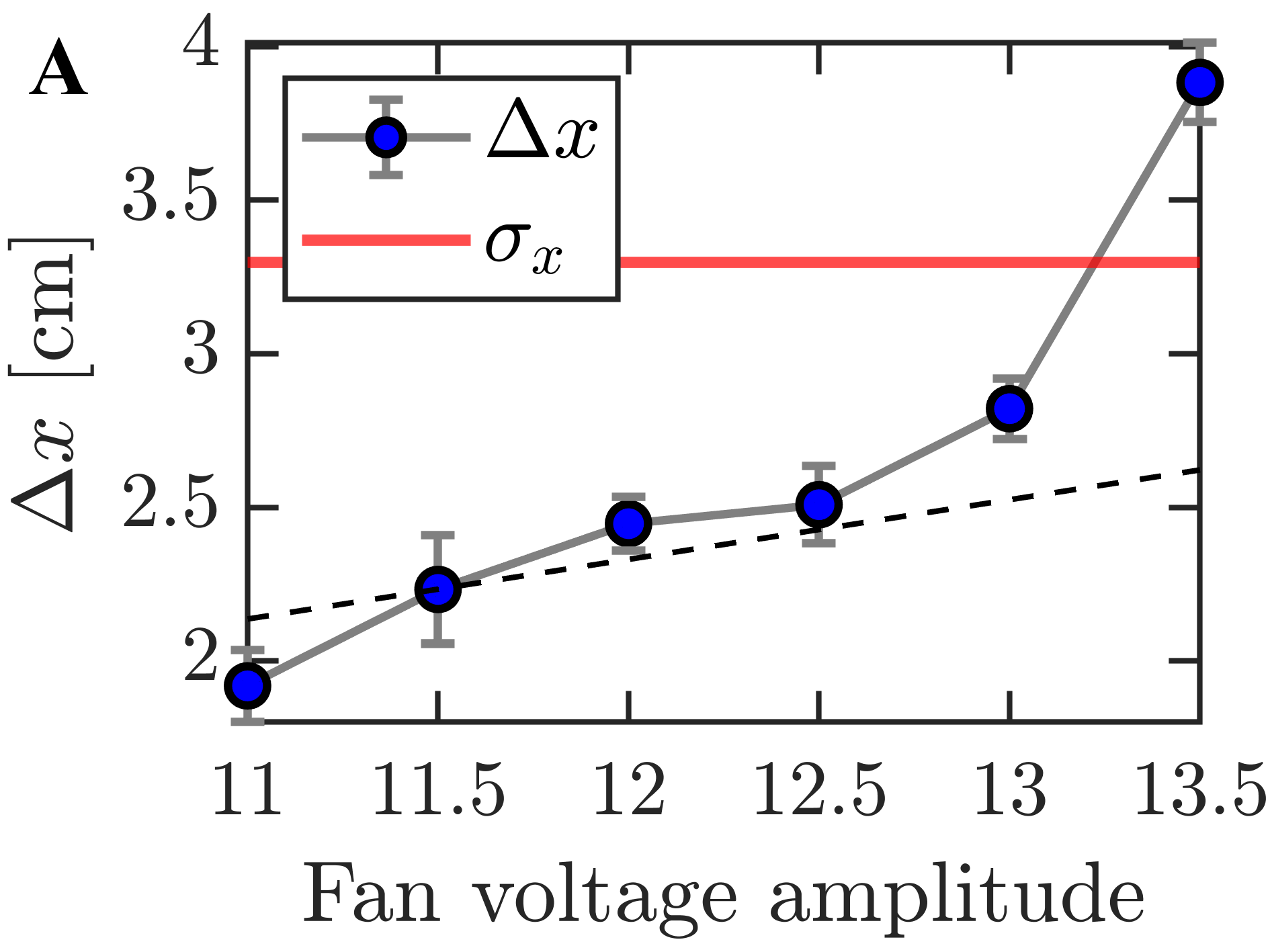}
    \includegraphics[width=0.20\textwidth]{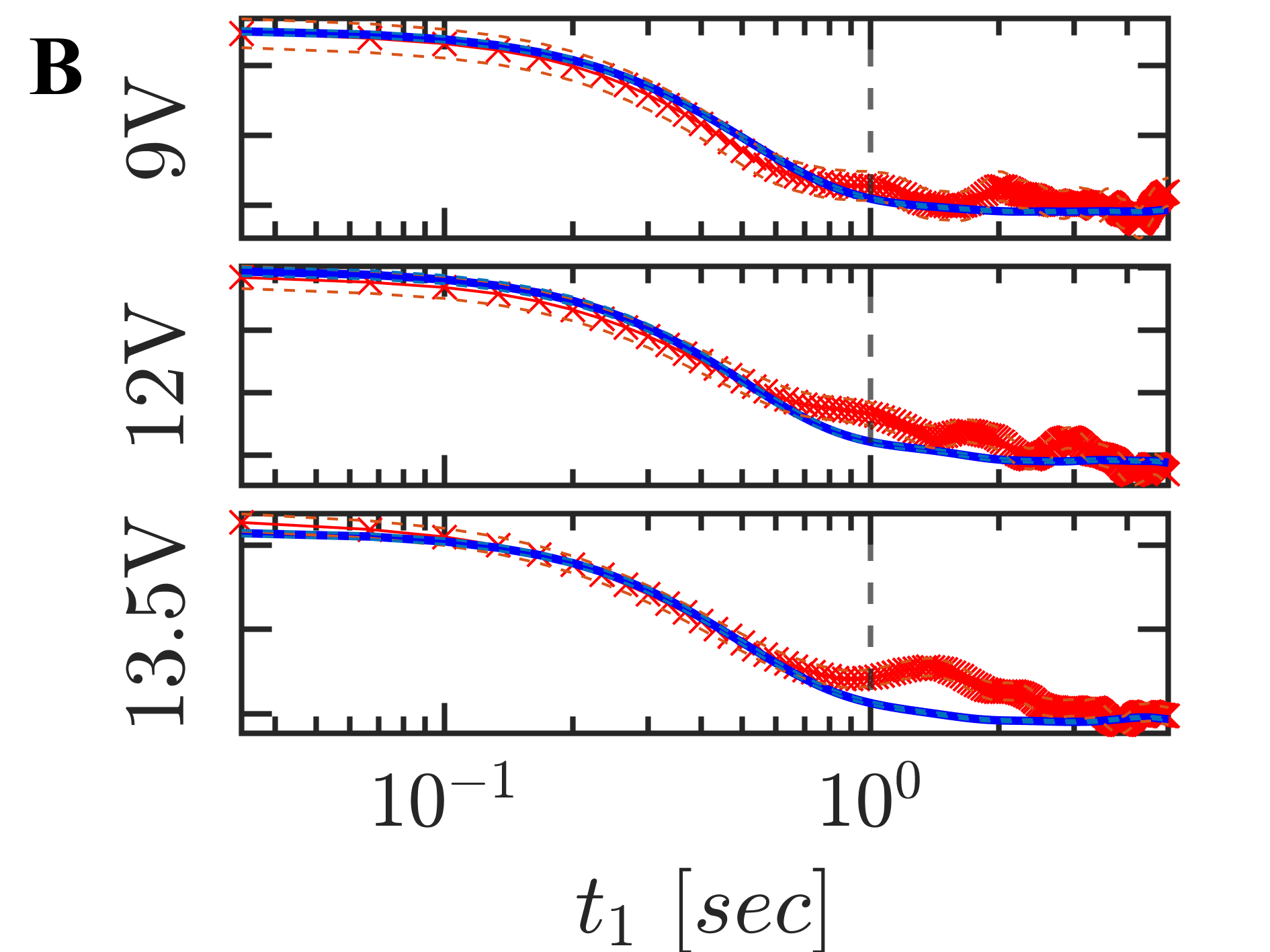}
    \caption{ 
    \textbf{The generalized FDR under different perturbation amplitudes.} \textbf{A.} \ The mean displacement $\Delta x$ under a perturbation of a tracer in a $N_b=10$ bbot bath, as a function of fan operating voltage $V$ (error bars are standard errors). 
    The applied force $F_0=k\Delta x$ increases above $V=11$V. The tracer's position standard deviation, $\sigma_x$, is used to define the range of small perturbations (solid line).
    The dashed line is a linear fit of the first 4 data points, with $\Delta x = 0.19\cdot V$, whereas $V=13$V and $13.5$V are out of the linear regime.
    \textbf{B.} \ The generalized FDRs with $T_{\text{eff}}\sim\langle \Delta x^2\rangle_0$ are plotted for $V=9, \ 12, \ 13.5$~V, with $N_b=10$. The results present an average over $M=375$ perturbation sequences.}
    \label{fig:SIfig6}
\end{figure}

\begin{figure}[h] 
    \centering
    \includegraphics[width=0.23\textwidth]{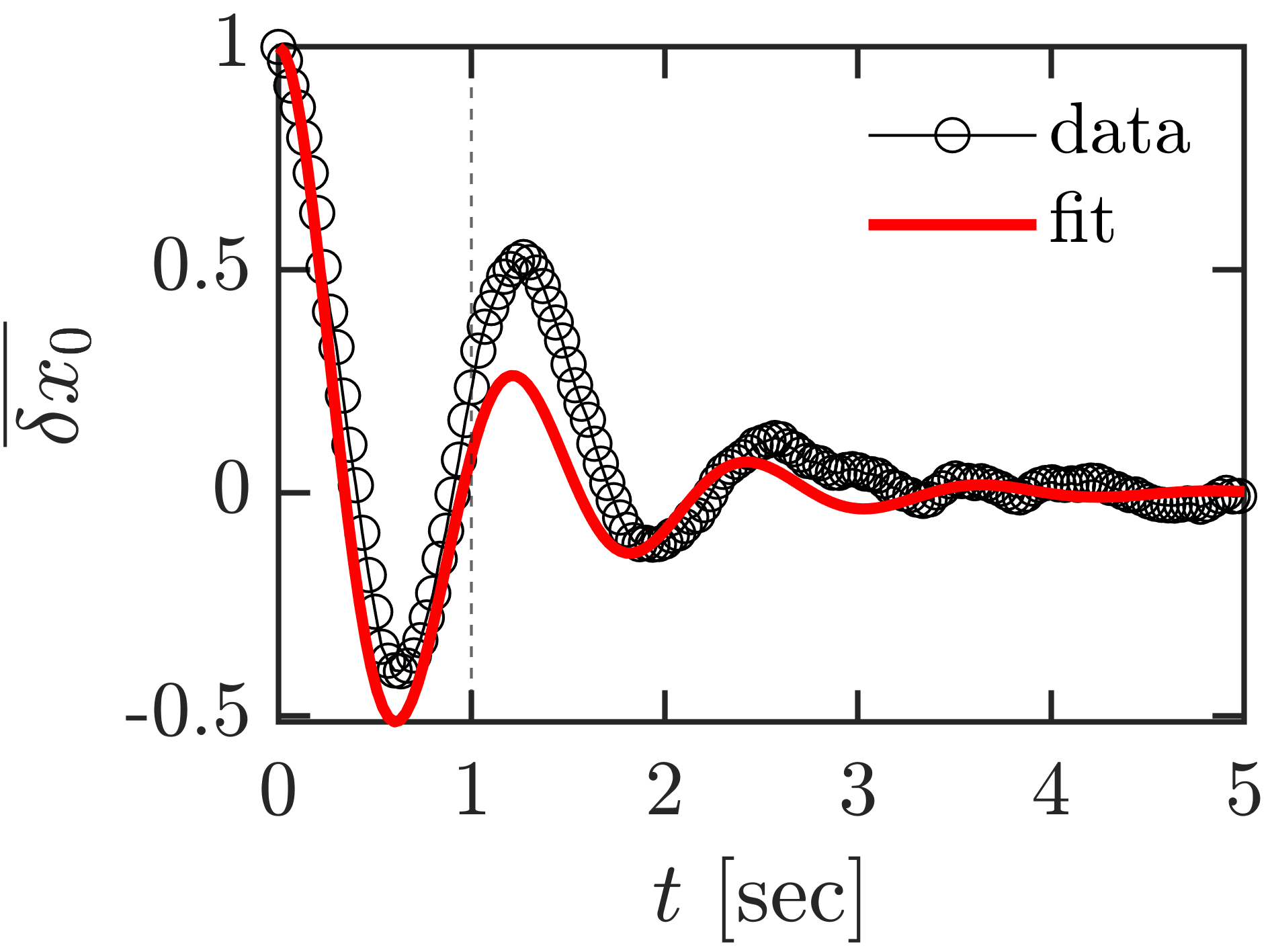}
    \caption{ 
    \textbf{Passive response measurement.}
    We probe the tracer's coupling to the trap in the absence of collisions by measuring the (deterministic) response $\delta x_0$ to a step-perturbation with the bbots removed ($13.5$V fan voltage, abruptly switched off at $t=0$). Considering the tracer relaxes as a damped harmonic oscillator with natural frequency $(\tau^0_\Omega)^{-1}=\sqrt{k/m}=5.3 \ \textrm{s}^{-1}$, the passive damping timescale is obtained by fitting the normalized response to Eq. 2, $\tau^0_r=0.9$s.
    }
    \label{fig:SIfig7}
\end{figure}

\begin{figure}[H] 
    \centering
    \includegraphics[width=0.20\textwidth]{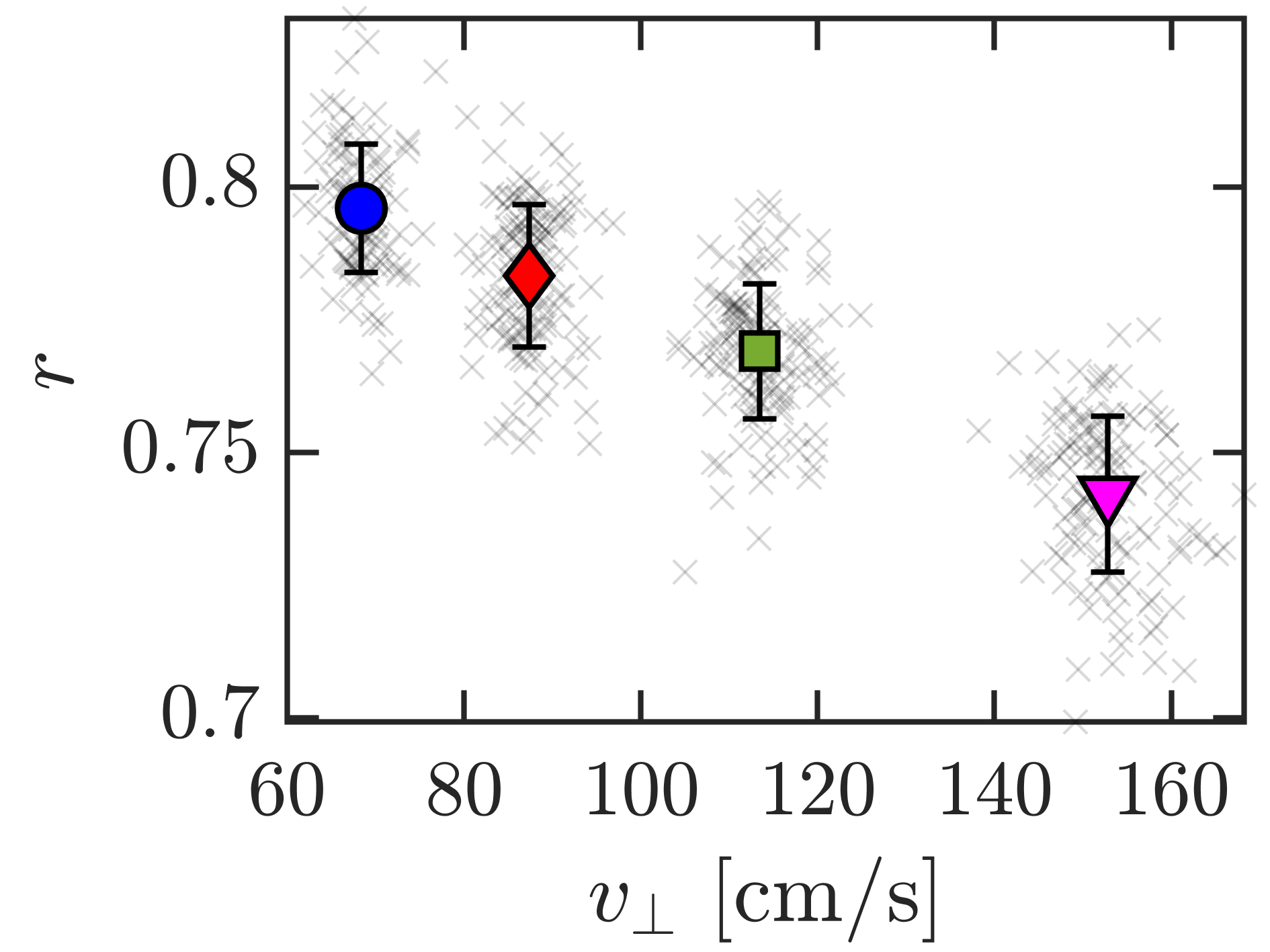}
    \caption{
    \textbf{Tracer's coefficient of restitution versus impact velocities.}
    The restitution coefficient $r$ was determined from the sound signal emitted by the styrofoam ball (the tracer) bouncing repeatedly off a surface, as detailed in Ref. \cite{heckel2016}.
    We obtain $r$ by measuring the time lag between consecutive impacts:
    $r_i = |v'_i/v_i|=(t_{i+1}-t_i)/(t_i-t_{i-1})$, for an impact velocity $v_i = \frac{g}{2}(t_{i-1}-t_{i})$, $i=2,3,...$. 
    Here, $v_i$ and $v'_i$ are the normal pre- and post-collision velocities occurring at time $t_i$, and $g=9.81 \textrm{m/s}^2$ is the gravitational acceleration. 
    The plot shows four successive impacts of a repeated experiment (50 trials), with mean values from last to first respectively: $r=0.79$ (circle), $0.78$ (diamond), $0.76$ (square), $0.74$ (triangle), for $v_{\perp}=68.37$, $87.35$, $113.4$, $152.83$ cm/s. 
    }
    \label{fig:SIfig8}
\end{figure}

\begin{figure}[H] 
    \centering
    \includegraphics[width=0.21\textwidth]{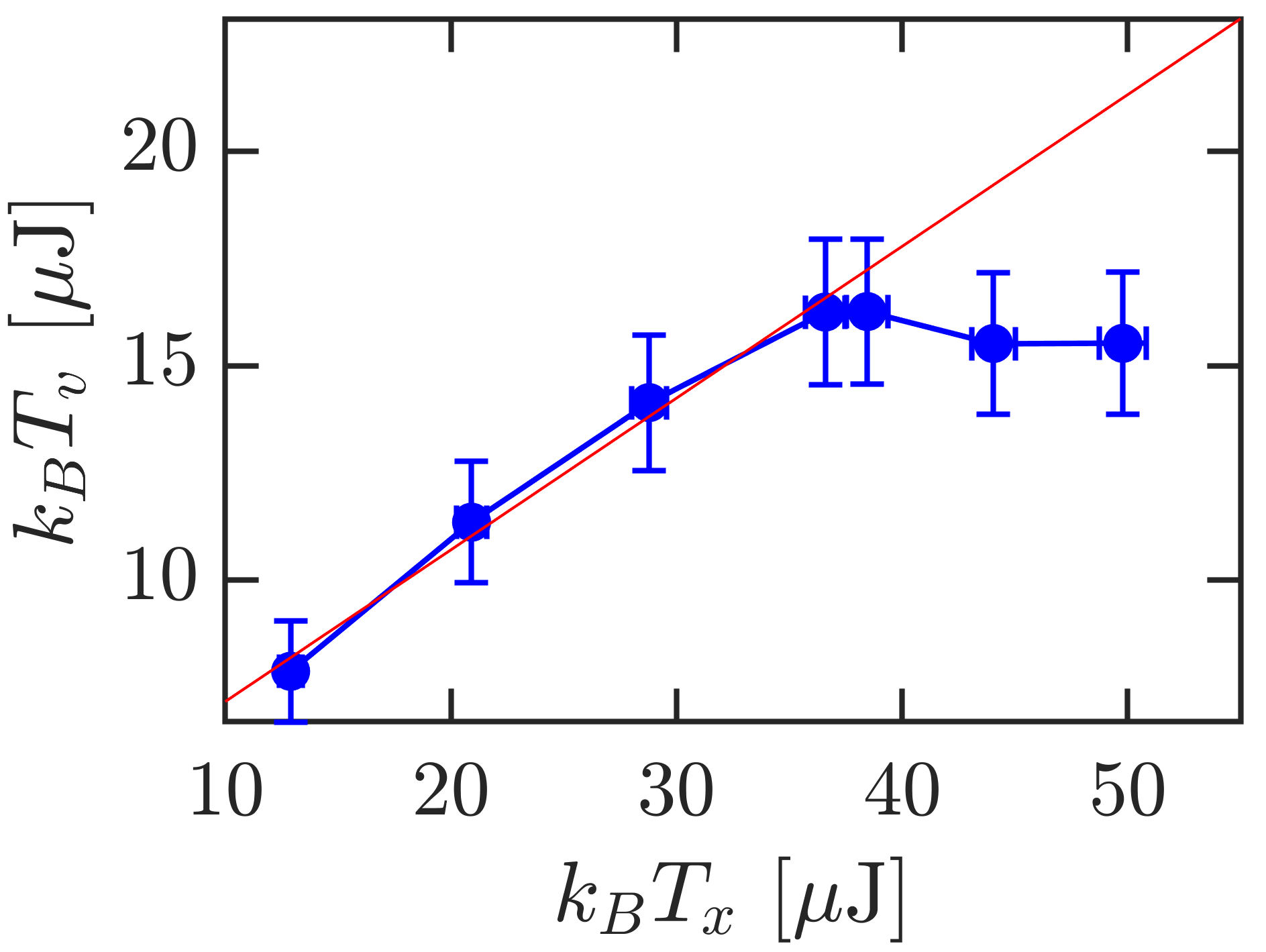}
    \caption{
    \textbf{Kinetic and potential temperatures.} 
    The potential temperature, $k_BT_{x} = k\langle \Delta x^2\rangle_0$, is plotted versus the kinetic temperature of the tracer, $k_BT_{v} = m\langle v^2\rangle_0$. We find a linear relation for the range of $3\geq N_b \leq10$. 
    The red line is the best linear fit for this range. 
    }
    \label{fig:SIfig10}
\end{figure}

\begin{figure}[H] 
    \centering
    \includegraphics[width=0.23\textwidth]{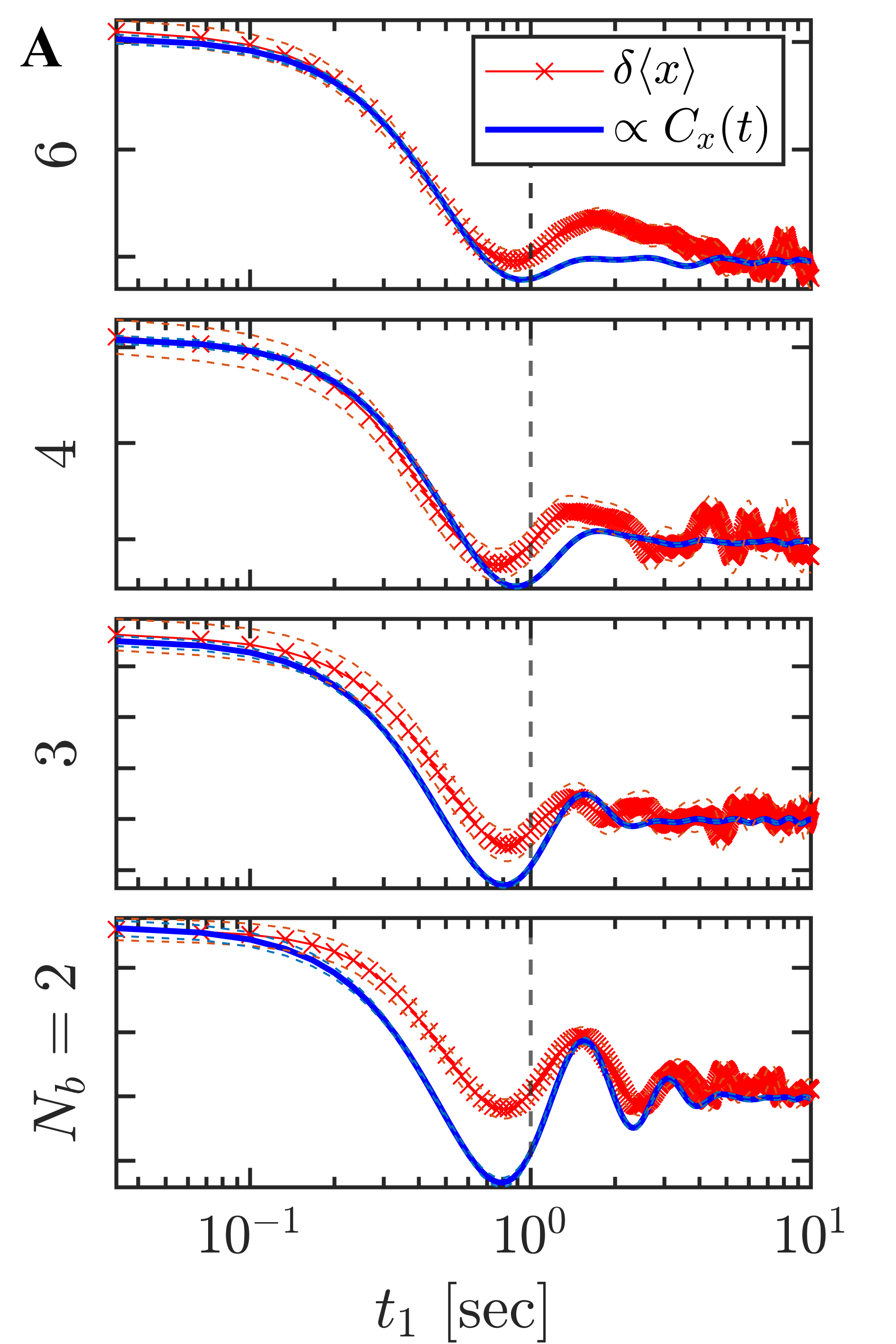}
    \includegraphics[width=0.23\textwidth]{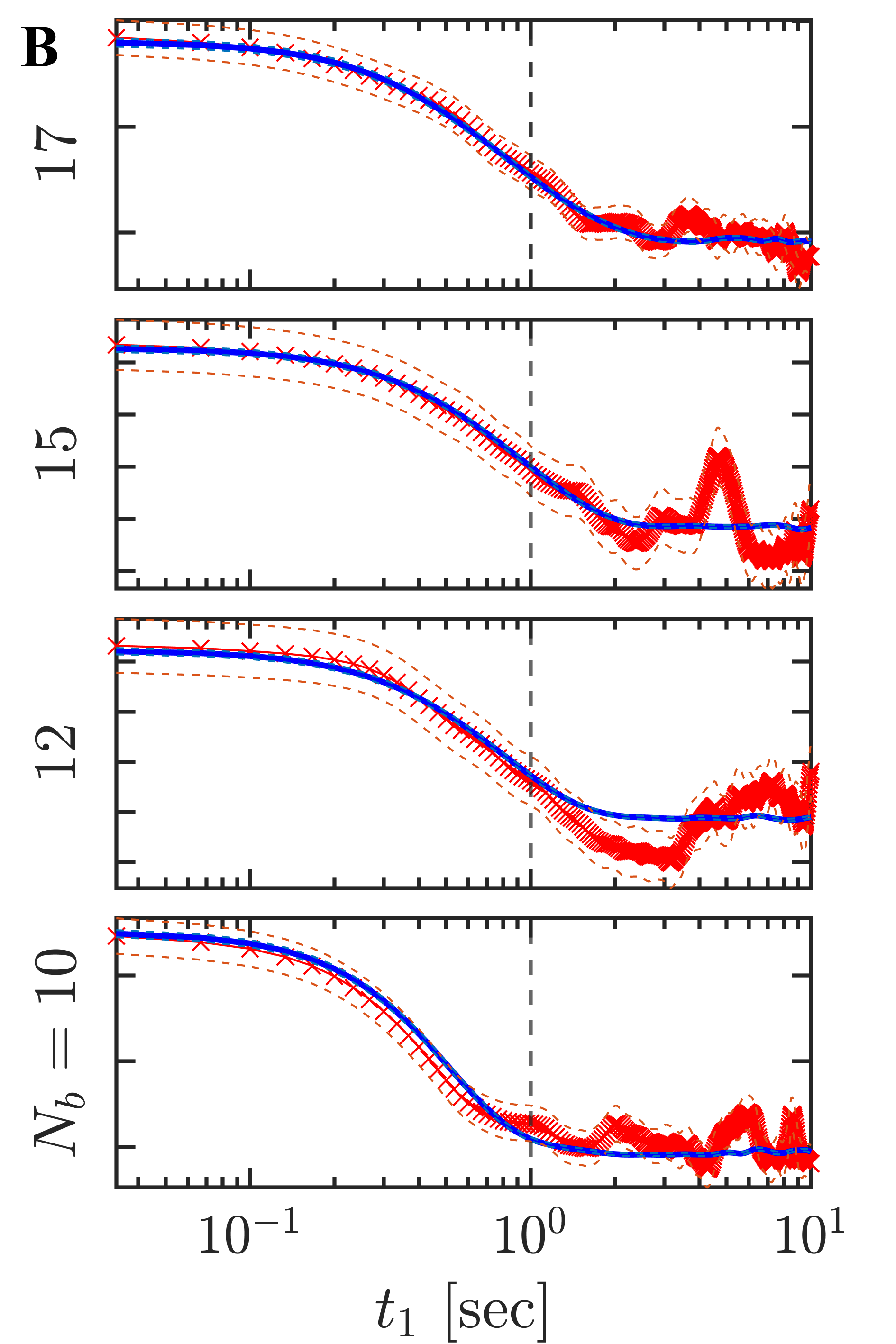}
    \caption{
    \textbf{FDR results with different numbers of bbots.} 
    \ \textbf{A,B.} Generalized FDRs with $T_{\text{eff}}\sim\langle \Delta x^2\rangle_0$ for $N_b = \{2,3,4,6,10,12,15,17\}$ bbot baths.
    The results were obtained for an ensemble average of $M=375$ perturbation sequences, with the same tracer $m\approx 1$~g, gravitational stiffness $k\approx 28.2~\text{g/s}^{-2}$, and fan operating voltage with $\Delta x(V)<\sigma_x$.
    The vertical dashed lines are $T_c=1$~s.
    }
    \label{fig:SIfig9}
\end{figure}

We furthermore add the following supplemental results:
A FDR test with constant $N_b$ and increasing perturbation amplitudes (Fig.~\ref{fig:SIfig6}); \
The passive response of the tracer to a step-perturbation in the absence of bbots (Fig.~\ref{fig:SIfig7}); \
The tracer's restitution coefficient measurement (Fig.~\ref{fig:SIfig8}); \
$T_{\text{eff}}$ versus the kinetic temperature $T_v\sim \langle v^2\rangle$ (Fig.~\ref{fig:SIfig10}); \
Individual FDR results for all $N_b$ (Fig.~\ref{fig:SIfig9}). 


\end{document}